\documentclass[lettersize,journal]{IEEEtran}
\usepackage{amsmath,amsfonts}
\usepackage{amssymb}
\usepackage{algorithmic}
\usepackage{algorithm}
\usepackage{array}
\usepackage[caption=false,font=normalsize,labelfont=normalsize,textfont=sf]{subfig}
\usepackage{textcomp}
\usepackage{stfloats}
\usepackage{url}
\usepackage{verbatim}
\usepackage{graphicx}
\usepackage{cite}
\usepackage{tikz}
\usetikzlibrary{calc}
\usetikzlibrary{backgrounds}

\colorlet{commentColor}{black}
\usepackage{breqn}
\usepackage{multicol}
\usepackage{mathtools}
\usepackage{siunitx}
\usepackage{xfrac}
\usepackage{nicefrac}

\usepackage{pgfplots}
\pgfplotsset{compat=1.18}
\usepgfplotslibrary{polar,fillbetween}

\colorlet{comment}{black}

\makeatletter
\newcommand{\pseudoeqref}[1]{\textup{\tagform@{#1}}}
\makeatother

\begin{document}

\title{On the Extension of Differential Beamforming Theory to\\Arbitrary Planar Arrays of First-Order Elements}
\author{Federico~Miotello,~\IEEEmembership{Member,~IEEE},
Davide~Albertini,~\IEEEmembership{Member,~IEEE},\\
Alberto~Bernardini,~\IEEEmembership{Senior Member,~IEEE}
\thanks{F. Miotello and A. Bernardini are with the Dipartimento di Elettronica, Informazione e Bioingegneria (DEIB), Politecnico di Milano, 20133 Milano, Italy (e-mail: federico.miotello@polimi.it; alberto.bernardini@polimi.it).}
\thanks{D. Albertini is with STMicroelectronics, 20864 Agrate Brianza (MB), Italy (e-mail: davide.albertini@st.com)}}



\maketitle

\begin{abstract}
Small-size acoustic arrays exploit spatial diversity to achieve capabilities beyond those of single-element devices, with applications ranging from teleconferencing to immersive multimedia. A key requirement for broadband array processing is a frequency-invariant spatial response, which ensures consistent directivity across wide bandwidths and prevents spectral coloration. \textcolor{comment}{Differential beamforming achieves such frequency-independent behavior} by leveraging pressure differences between closely spaced elements in compact arrays. Traditional differential approaches, however, assume omnidirectional array elements, whereas real transducers often exhibit frequency-dependent directivity that can degrade performance if not properly modeled. To address this limitation, we propose a generalized modal-matching framework for frequency-invariant differential beamforming, applicable to arbitrary planar arrays of first-order directional elements. By representing the target beampattern as a truncated circular harmonic expansion and fitting it to the actual element responses, the proposed method accommodates arbitrary array geometries, element directivities, and orientations. The resulting framework enables the synthesis of high-order steerable beampatterns while avoiding restrictive layout assumptions. Simulations confirm that explicitly accounting for transducer directivity at the design stage yields accurate and robust performance across frequencies, array geometries, and noise conditions.
\end{abstract}

\begin{IEEEkeywords}
Differential beamforming, array signal processing, microphone arrays, loudspeaker arrays, small-size arrays, first-order directivity, frequency-invariant beampattern.
\end{IEEEkeywords}

\section{Introduction}
\label{sec:01_intro}

\IEEEPARstart{I}{n recent years}, research on compact acoustic arrays has emerged as a trending topic of interest in audio signal processing, largely due to the widespread availability of small-size microphones and loudspeakers, such as MEMS-based devices \cite{zawawi2020review, garud2023mems, wang2021review}.
Through the spatial processing of acoustic signals (i.e., beamforming), compact arrays enable diverse applications that span various domains, including automotive engineering \cite{buck2001first, buck2009compact}, hands-free telecommunication \cite{elko1996microphone}, wearable devices \cite{levin2016near, feng2025directional, deppisch2024blind}, as well as tasks like speech enhancement and noise suppression for teleconferencing and immersive multimedia \cite{huang2025advances, benesty2018fundamentals, benesty2017fundamentals}.

In the literature, beamforming techniques for small-size arrays are typically grouped into two main categories: additive approaches \cite{kim2013sound, brandstein2001microphone, benesty2008microphone, huang2006acoustic, monzingo2004introduction} and differential approaches \cite{kim2013sound, elko2004differential, benesty2012study, benesty2016fundamentals, benesty2015design}.
In particular, differential beamforming \cite{benesty2016fundamentals} represents a prominent class of techniques that leverages small-size array elements to provide highly directional and almost frequency-invariant spatial responses.
In fact, in broadband acoustic processing, maintaining consistent spatial directivity across wide bandwidths is critical to avoid spectral coloration and ensure robust and reliable performance across diverse applications.
To this end, differential beamforming leverages pressure differentials computed through finite differences between closely spaced transducers \cite{benesty2016fundamentals}.
Thus, the effectiveness of this approach inherently relies on the proximity of array elements, which enables accurate approximation of pressure gradients and naturally yields frequency invariance as an emergent property.
Building on this principle, Differential Microphone Arrays (DMAs) exploit beamforming to enhance signals from desired directions while attenuating interference and noise from undesired ones \cite{elko2004differential,de2011design, benesty2012study}.
Conversely, in the reciprocal scenario, Differential Loudspeaker Arrays (DLAs) exploit beamforming for directional sound radiation to effectively focus acoustic energy toward specific spatial regions \cite{wang2021design, miotello2023steerable, zhang2022sound, kim2013sound}.

The literature is rich with differential beamforming techniques, based on various array geometries and characteristics. 
Building on the early work on pressure-gradient microphones \cite{weinberger1933uni, olson1946gradient} and gradient loudspeakers \cite{olson1973gradient, gudvangen2014properties}, differential array theory has been developed, with a focus on DMAs, both in the time domain \cite{elko1997steerable, teutsch2001first, elko2004differential, elko2008microphone, benesty2016fundamentals, buchris2019design}, which suits real-time applications with tight latency constraints \cite{borra2020efficient}, and in the short-time Fourier transform (STFT) domain, which allows for frequency-dependent weighting \cite{benesty2016fundamentals,benesty2012study,benesty2015design}. STFT-based DLAs have also been investigated in \cite{wang2021design, miotello2023steerable, zhang2022sound}.

Most existing designs are based on uniform linear \cite{chen2014design, benesty2016fundamentals, benesty2012study, Bernardini2018WDULADMA} or uniform circular arrays \cite{huang2017designCircular, benesty2015design, zhang2022sound, lovatello2018steerable}. Uniform linear arrays offer limited steering capabilities, as effective single-mainlobe beams are typically achievable only in end-fire directions, and are subject to front–back ambiguity \cite{borra2020efficient}. Uniform circular arrays, instead, provide improved steering in the array plane \cite{borra2020efficient}, but may suffer from the so-called deep nulls in both white noise gain and directivity factor \cite{huang2018design, wang2024design}, potentially resulting in beam distortion. To address these limitations, alternative geometries, such as nonuniform linear \cite{zhang2015studyNonuniform} and concentric circular arrays \cite{huang2017design, huang2018design, wang2024design}, have been proposed.

Linear and circular arrays, while attractive for their simple layout, impose strict geometry constraints. In many practical scenarios (e.g., compact embedded systems), a uniform element spacing or perfectly regular geometry may be infeasible due to design requirements, manufacturing tolerances, or element misplacement. These challenges motivate the development of beamforming methods capable of handling arbitrary array geometries \cite{borra2020efficient, albertini2022TwoStage, huang2018designArbitrary, huang2020steerable}. To this end, a particularly influential family of beamformers is the \emph{frequency-invariant beamforming via least-squares error} (FIB-LSE) \cite{huang2018designArbitrary, huang2017designCircular}. Applied to both irregular planar layouts and structured array configurations, FIB-LSE methods express the desired beampattern as a truncated modal (circular harmonic) expansion and match it to the array’s actual response via a least-squares fit, exploiting the well-known Jacobi–Anger expansion. This strategy decouples beampattern synthesis from array geometry and has been applied successfully to both microphone \cite{huang2017designCircular, huang2018designArbitrary, albertini2022TwoStage} and loudspeaker \cite{zhang2022sound} arrays, achieving high-order directivity with compact apertures and arbitrary layouts.
It is worth noting that such modal-matching formulations relate to the field of modal array signal processing \cite{teutsch2007modal}, in which spatially distributed measurements are used to decompose an acoustic field into orthogonal eigenbeams.
\textcolor{comment}{The conceptual similarity between differential beamforming and modal beamforming has been explicitly discussed in \cite{huang2017design}, where modal beamformers, and in particular circular harmonic beamformers \cite{tiana2010beamforming, tiana2011beamforming, torres2012robust, yan2015optimal}, are interpreted within the broader FIB-LSE framework.
Moreover, although the present work focuses on 2D arbitrary geometries, similar modal approaches also apply to spherical microphone arrays \cite{rafaely2015fundamentals, iijima2021binaural, abhayapala2002theory, jarrett2017theory, hacihabibouglu2024spherically}, further highlighting a shared foundation between differential and modal beamforming.}

Beyond element positioning, the inherent directivity of the array elements also plays a critical role in beamformer design.
Although directional elements have been considered in Ambisonics and spherical microphone array literature \cite{rafaely2015fundamentals, zotter2019ambisonics, iijima2021binaural}, most existing differential designs, including the original FIB-LSE formulation \cite{huang2017designCircular, huang2018designArbitrary}, assume that each transducer is an ideal omnidirectional sensor (or radiator).
\textcolor{comment}{In reality, small-size microphones and loudspeakers often exhibit intrinsic frequency-dependent directivity that can distort the intended spatial response if not properly accounted for.
Incorporating these element-wise characteristics into the beamformer formulation requires not only accurate modeling, but also a generalized filter design framework capable of consistently embedding such directional responses into the modal-matching process, enabling accurate, broadband, and robust beamforming.}
To this end, recent work has begun to account for the directional behavior of transducers in differential arrays. 
Luo et al.\ \cite{luo2023design} and Miotello et al.\ \cite{miotello2023steerable} consider linear differential arrays composed of alternating omnidirectional and dipole elements, while Luo et al.\ \cite{luo2024design} and Huang et al.\ \cite{huang2025robust} extend this concept to arrays of omnidirectional and directional elements with uniform or non-uniform look directions.
Wang et al.\ \cite{wang2024design} extend the framework in \cite{huang2018design} by considering uniform concentric circular arrays of first-order elements steered radially outward. In particular, \cite{wang2024design} took an important step beyond Jacobi–Anger–based approaches by formulating the problem directly through a circular harmonic decomposition of the element-wise responses.
Nonetheless, although these methods incorporate, to some extent, directional elements into conventional differential arrays, they remain restricted to specific geometries (linear or circular), and also impose strict constraints on the inherent directivity of the array elements, thereby limiting their applicability in more flexible or irregular configurations.

To address the aforementioned limitations, this paper introduces a generalized framework for differential beamforming that accommodates arbitrary planar arrays of first-order directional elements.
The proposed approach enables the synthesis of high-order steerable beampatterns without imposing constraints on array-element positions, beam shapes, or steering directions.
By decoupling array geometry from beampattern design through modal matching, the framework extends existing differential beamforming methods, in particular \cite{huang2018designArbitrary} and \cite{wang2024design}, which emerge as specific cases within the proposed formulation.
It is important to note that, due to the reciprocal nature of acoustic transduction, the filter design technique presented in this manuscript applies equally to both DMAs and DLAs \cite{elko2004differential}.
For this reason, although we will treat the array elements as microphones, the proposed derivations are general enough to apply to the reciprocal loudspeaker scenario.

The rest of the manuscript is organized as follows.
Sec.~\ref{sec:02_background} reviews the background theory on differential beamforming for planar arrays of omnidirectional elements, with a focus on the FIB-LSE design proposed in \cite{huang2018designArbitrary}. In Sec.~\ref{sec:03_method}, we extend this design to accommodate arbitrary planar arrays of first-order elements. Sec.~\ref{sec:04_validation} presents simulations and experiments to validate the proposed framework. Finally, Sec.~\ref{sec:05_conclusion} draws conclusions.
\section{Background on Planar Differential Arrays of Omnidirectional Elements}
\label{sec:02_background}
\subsection{Signal Model}\label{sec:signal_model}
Let us consider an array composed of $M$ microphones arbitrarily placed in a two‐dimensional plane.
Considering a Cartesian coordinate system, the position of the $m$th array element can be expressed as:
\begin{equation}\label{eq:positions}
    \mathbf{r}_m = r_m\left[ \cos{(\phi_m)}, \sin{(\phi_m)} \right],
\end{equation}
where $r_m$ is its distance from the origin of the axes and $\phi_m$ is the angle it forms with respect to the $x$-axis.
Considering a distant source, we can express the signal captured by the $m$th microphone as
\begin{equation}\label{eq:captured_signal}
    Y_m(\omega) = D_m(\omega, \theta_{\mathrm{s}})X(\omega) + V_m(\omega),
\end{equation}
where $\omega = 2\pi f$ is the angular frequency, $f$ is the temporal frequency, $X(\omega)$ is the source signal, $V_m(\omega)$ models additive noise and $D_m(\omega, \theta_{\mathrm{s}})$ \textcolor{comment}{models the acoustic wave propagation from the source to the $m$th microphone}, with $\theta_{\mathrm{s}}$ being the direction of arrival of the source.
Under far-field conditions, $D_m(\omega, \theta_{\mathrm{s}})$ can be expressed as
\begin{equation}\label{eq:steering_vector}
    D_m(\omega, \theta_{\mathrm{s}}) = e^{jk r_m\cos(\theta_{\mathrm{s}} - \phi_m)},
\end{equation}
where $k = \frac{\omega}{c}$ is the wave number, $c$ is the speed of sound in air and $j$ is the imaginary unit.

In matrix notation, \eqref{eq:captured_signal} can be rewritten as
\begin{equation}\label{eq:captured_signal_matrix}
    \mathbf{y}(\omega) = \mathbf{d}(\omega, \theta_{\mathrm{s}})X(\omega) + \mathbf{v}(\omega),
\end{equation}
in which we have
\begin{equation}\label{eq:vectors}
    \begin{aligned}
        \mathbf{y}(\omega) &= [Y_1(\omega), \dots, Y_M(\omega)]^T,\\
        \mathbf{d}(\omega, \theta_{\mathrm{s}}) &= [D_1(\omega, \theta_{\mathrm{s}}), \dots, D_M(\omega, \theta_{\mathrm{s}})]^T,\\
        \mathbf{v}(\omega) &= [V_1(\omega), \dots, V_M(\omega)]^T.
    \end{aligned}
\end{equation}
Spatial filtering is performed by combining microphone signals $Y_m(\omega)$ as follows
\begin{equation}\label{eq:filtering}
    Z(\omega) = \sum_{m=1}^M H_m^*(\omega)Y_m(\omega) = \mathbf{h}^H(\omega)\mathbf{y}(\omega),
\end{equation}
where $H_m(\omega)$ is the complex filter coefficient applied to the $m$th microphone signal, the asterisk $^*$ denotes complex conjugation, $Z(\omega)$ is an estimate of $X(\omega)$, the superscript $^H$ denotes the Hermitian operator, and $\mathbf{h}(\omega) = [H_1(\omega), \dots, H_M(\omega)]^T$ is the beamforming filter or beamformer.

\subsection{Metrics}\label{sec:metrics}
The three key performance metrics commonly used in the beamforming literature to evaluate the effectiveness of the beamformer $\mathbf{h}(\omega)$ are the beampattern, the white noise gain, and the directivity factor. In particular, the latter two correspond to specific forms of signal-to-noise ratio gain with respect to temporally and spatially white noise and diffuse noise, respectively \cite{benesty2012study, borra2020efficient}.
\subsubsection{Beampattern}
The sensitivity, or spatial response, of a beamforming filter $\mathbf{h}(\omega)$ to a plane wave impinging on the array with an angle $\theta$ is referred to as beampattern, and is formally defined as
\begin{equation}\label{eq:beampattern}
\begin{aligned}
    \mathcal{B}[\mathbf{h}(\omega),\theta] = \mathbf{h}^H(\omega)\mathbf{d}(\omega, \theta).
\end{aligned}
\end{equation}
When dealing with data-independent beamformers, in which the filter $\mathbf{h}(\omega)$ is designed to perform beam steering in direction $\theta_{\mathrm{s}}$, the function $|\mathcal{B}[\mathbf{h}(\omega),\theta]|^2$ exhibits a maximum when evaluated for $\theta = \theta_{\mathrm{s}}$ \cite{borra2020efficient}.
\subsubsection{White Noise Gain}
The white noise is defined as
\begin{equation}\label{eq:wng}
    \mathcal{W}[\mathbf{h}(\omega)] = \frac{|\mathcal{B}[\mathbf{h}(\omega),\theta_{\mathrm{s}}]|^2}{\mathbf{h}^H(\omega)\mathbf{h}(\omega)}.
\end{equation}
It quantifies the sensitivity of the beamformer $\mathbf{h}(\omega)$ to sensor imperfections, such as self-noise, gain or phase mismatches, and element misplacement, with larger values indicating greater robustness \cite{tu2022theoretical, wang2024design}.
\subsubsection{Directivity Factor}
The directivity is defined as \cite{wang2024design}
\begin{equation}\label{eq:df}
    \mathcal{D}[\mathbf{h}(\omega)] = \frac{|\mathcal{B}[\mathbf{h}(\omega),\theta_{\mathrm{s}}]|^2}{\frac{1}{2\pi}\int_{-\pi}^\pi |\mathcal{B}[\mathbf{h}(\omega),\theta]|^2\,d\theta}.
\end{equation}
It quantifies the ability of the beamformer $\mathbf{h}(\omega)$ to enhance signals from the steering direction $\theta_{\mathrm{s}}$ relative to diffuse noise.

\subsection{Filter Design}\label{sec:filter_design}
A frequency-invariant beampattern that is symmetric with respect to the steering direction $\theta_{\mathrm{s}}$ can be expressed as a weighted sum of complex sinusoids as
\begin{equation}\label{eq:target_beampattern}
    \begin{aligned}
        \bar{\mathcal{B}}(\mathbf{b}, \theta, \theta_{\mathrm{s}}) = \sum_{n=-N}^N b_n e^{jn(\theta - \theta_{\mathrm{s}})} = \mathbf{b}^T\mathbf{\Upsilon}(\theta_{\mathrm{s}})\mathbf{c}(\theta),
    \end{aligned}
\end{equation}
where $N$ is the order of the beampattern,
and
\begin{equation}
    \begin{aligned}
        \mathbf{b} &= [b_{-N}, \dots, b_0, \dots, b_N]^T,\\
        \mathbf{\Upsilon}(\theta_{\mathrm{s}}) &= \operatorname{diag}(e^{jN\theta_{\mathrm{s}}}, \dots, 1, \dots, e^{-jN\theta_{\mathrm{s}}}),\\
        \mathbf{c}(\theta) &= [e^{-jN\theta}, \dots, 1, \dots, e^{jN\theta}]^T,
    \end{aligned}
\end{equation}
where $\operatorname{diag}(\cdot)$ denotes a square diagonal matrix with the given elements on the main diagonal.
In particular, $\mathbf{b}$ is a column vector of $2N+1$ real coefficients that determine the shape of the beampattern, $\mathbf{c}(\theta)$ is a vector of circular harmonics, and $\mathbf{\Upsilon}(\theta_{\mathrm{s}})$ is a rotation matrix that steers the beampattern toward $\theta_{\mathrm{s}}$.
For further reading on the design of target beampatterns, the reader is referred to \cite{huang2017designCircular, huang2018designArbitrary, benesty2016fundamentals, huang2020steerable, benesty2015design}.
The coefficients of widely adopted directivity patterns, which can be derived under various optimization criteria, can be found in \cite{de2011design, huang2020steerable}.

In many differential beamforming approaches \cite{huang2017designCircular, huang2018designArbitrary, borra2020efficient}, the filter $\mathbf{h}(\omega)$ is designed in such a way that the resulting beampattern described in \eqref{eq:beampattern} matches a target beampattern in the form presented in \eqref{eq:target_beampattern}.
Moreover, to prevent any amplitude modification of the signals coming from direction $\theta_{\mathrm{s}}$, it is common to impose that the target beampattern satisfies the so-called distortionless constraint \cite{benesty2015design}, i.e.,  $\bar{\mathcal{B}}(\mathbf{b}, \theta_{\mathrm{s}}, \theta_{\mathrm{s}}) = 1$.

By imposing $\mathcal{B}[\mathbf{h}(\omega),\theta] = \bar{\mathcal{B}}(\mathbf{b}, \theta, \theta_{\mathrm{s}})$ we get
\begin{equation}\label{eq:renderedbp-targetbp}
    \begin{aligned}
        \sum_{m=1}^M H_m^*(\omega)e^{jkr_m\cos(\theta - \phi_m)} = \sum_{n=-N}^N b_n e^{jn(\theta - \theta_{\mathrm{s}})},
    \end{aligned}
\end{equation}
which can be rewritten by exploiting the well-known Jacobi-Anger expansion \cite{cuyt2008handbook} defined as
\begin{equation}\label{eq:ja_expansion}
e^{jkr_m\cos(\theta - \phi_m)} = \sum_{n=-\infty}^{+\infty}j^nJ_n(kr_m)e^{jn(\theta-\phi_m)},
\end{equation}
where $J_n(\cdot)$ is the $n$th order Bessel function of the first kind.
After truncating the expansion to the order $N$, it is possible to rewrite \eqref{eq:renderedbp-targetbp} as
\begin{equation}\label{eq:renderedbp-targetbp-ja}
    \begin{aligned}
        \sum_{m=1}^M H_m^*(\omega)\sum_{n=-N}^{N}j^nJ_n(kr_m)e^{jn(\theta-\phi_m)}
        =\sum_{n=-N}^N b_n e^{jn(\theta - \theta_{\mathrm{s}})}.
    \end{aligned}
\end{equation}
The dependence on $\theta$ can be eliminated from both sides of \eqref{eq:renderedbp-targetbp-ja}, thus enabling a mode-by-mode matching which yields
\begin{equation}\label{eq:modal-matching}
\sum_{m=1}^M H_m^*(\omega)j^nJ_n(kr_m)e^{-jn\phi_m} = b_n e^{-jn\theta_{\mathrm{s}}}.
\end{equation}
For each $n \in \{ -N, \dots, N\}$, \eqref{eq:modal-matching} can be expressed in matrix notation as
\begin{equation}\label{eq:modal_matching_omni}
 \mathbf{\Psi}(\omega)\mathbf{h}^*(\omega) = \mathbf{\Upsilon}(\theta_{\mathrm{s}})\mathbf{b},
\end{equation}
where
\begin{equation}\label{eq:psi_matrix}
    \mathbf{\Psi}(\omega) = \begin{bmatrix}
        j^{-N}\boldsymbol{\psi}^H_{-N}(\omega) \\
        \vdots\\
        \boldsymbol{\psi}^H_{0}(\omega)\\
        \vdots\\
        j^{N}\boldsymbol{\psi}^H_{N}(\omega)
    \end{bmatrix}
\end{equation}
is a $(2N+1) \times M$ matrix and
\begin{equation}
    \boldsymbol{\psi}_n(\omega) = \left[ J_n(kr_1)e^{jn\phi_1}, \dots, J_n(kr_m)e^{jn\phi_M} \right]^T.
\end{equation}
After conjugating both sides of \eqref{eq:modal_matching_omni} and obtaining $\mathbf{\Psi}^*(\omega)\mathbf{h}(\omega) = \mathbf{\Upsilon}^*(\theta_{\mathrm{s}})\mathbf{b}$, assuming $M \geq 2N+1$ and $\mathbf{\Psi}(\omega)$ has full row rank, the beamforming filter $\mathbf{h}(\omega)$ can be obtained as the minimum-norm solution 
\begin{equation}
    \mathbf{h}(\omega) = [\mathbf{\Psi}^*(\omega)]^\dagger\mathbf{\Upsilon}^*(\theta_{\mathrm{s}})\mathbf{b},
\end{equation}
where $[\mathbf{A}]^\dagger = [\mathbf{A}^H\mathbf{A}]^{-1}\mathbf{A}^H$ indicates the pseudoinverse of matrix $\mathbf{A}$.
\section{Planar Differential Arrays of First-Order Elements}
\label{sec:03_method}

\subsection{Signal Model}
We now extend the framework presented in Sec.~\ref{sec:signal_model} to array configurations in which each element $m$ exhibits a first-order symmetric directivity pattern, defined as
\begin{equation}\label{eq:transfer_function}
    T_m(\omega, \theta)=\left[1-q_m(\omega)\right]+q_m(\omega)\cos{(\theta-\theta_m)}, 
\end{equation}
where $0 \leq q_m(\omega) \leq 1$ is a frequency-dependent parameter controlling the element beam shape at frequency $\omega$, and $\theta_m$ is the element steering angle. The coefficient $q_m(\omega)$ determines the relative contribution of the omnidirectional component and the steered dipole component of the first-order directional pattern: if $q_m(\omega)=0$ then the pattern is omnidirectional and presents a spatially isotropic response, while if $q_m(\omega)=1$ the $m$th element is a dipole steered toward angle $\theta_m$. For real transducers, the parameter $q_m(\omega)$ could be obtained either from the manufacturer's datasheet or experimentally by measuring the angular response of each element and fitting it to the first-order model in \eqref{eq:transfer_function}.
Compact measurement setups, such as the one in \cite{herzog2025mems}, allow sampling $T_m(\theta)$ at high angular resolution, from which $q_m(\omega)$ and $\theta_m$ can be robustly estimated.

\begin{figure}[tbp]
    \centering
    \begin{tikzpicture}[scale=1.8]

\coordinate (O) at (0,0);

\def\psiOne{55}
\def\psiTwo{210}
\def\psiThree{120}
\def\psiM{320}

\def\rOne{0.7}
\def\rTwo{1.3}
\def\rThree{1.8}
\def\rL{1.6}

\coordinate (r1Coord) at (\psiOne:\rOne);
\coordinate (r2Coord) at (\psiTwo:\rTwo);
\coordinate (r3Coord) at (\psiThree:\rThree);
\coordinate (rLCoord) at (\psiM:\rL);

\draw[->] (-1.8,0) -- (1.8,0)   node[right] {$x$};
\draw[->] (0,-1.6) -- (0,1.8)   node[above] {$y$};

\coordinate (arc1Start) at (0:\rOne-0.2);
\coordinate (arc1End)   at (\psiOne:\rOne-0.2);
\coordinate (arc2Start) at (0:\rTwo-0.2);
\coordinate (arc2End)   at (\psiTwo:\rTwo-0.2);
\coordinate (arc3Start) at (0:\rThree-0.2);
\coordinate (arc3End)   at (\psiThree:\rThree-0.2);
\coordinate (arcMStart) at (0:\rL-0.27);
\coordinate (arcMEnd)   at (\psiM:\rL-0.27);

\draw[dashed,->,gray] (arc1Start) arc[start angle=0, end angle=\psiOne, radius=\rOne-0.2];
\draw[dashed,->,gray] (arc2Start) arc[start angle=0, end angle=\psiTwo, radius=\rTwo-0.2];
\draw[dashed,->,gray] (arc3Start) arc[start angle=0, end angle=\psiThree, radius=\rThree-0.2];
\draw[dashed,->,gray] (arcMStart) arc[start angle=0, end angle=\psiM,   radius=\rL-0.27];

\draw[solid,gray] (O) -- (r1Coord)
  node[pos={(\rOne - 0.2)/(2*\rOne)}, fill=white, inner sep=1pt] {\textcolor{black}{$r_1$}};

\draw[solid,gray] (O) -- (r2Coord)
  node[pos={(\rTwo - 0.2)/(2*\rTwo)}, fill=white, inner sep=1pt] {\textcolor{black}{$r_3$}};

\draw[solid,gray] (O) -- (r3Coord)
  node[pos={(\rThree - 0.2)/(2*\rThree)}, fill=white, inner sep=3pt] {\textcolor{black}{$r_2$}};

\draw[solid,gray] (O) -- (rLCoord)
  node[pos={(\rL - 0.2)/(2*\rL)}, fill=white, inner sep=2pt] {\textcolor{black}{$r_M$}};

\begin{scope}[shift=(r1Coord), rotate=135]
  \draw[fill]
    plot[domain=0:360, samples=360, variable=\t]
      ({0.2*abs(0.3+0.7*cos(\t))*cos(\t)},
       {0.2*abs(0.3+0.7*cos(\t))*sin(\t)})
    -- cycle;
\end{scope}

\begin{scope}[shift=(r2Coord), rotate=212]
  \draw[fill]
    plot[domain=0:360, samples=360, variable=\t]
      ({0.2*abs(0.3+0.7*cos(\t))*cos(\t)},
       {0.2*abs(0.3+0.7*cos(\t))*sin(\t)})
    -- cycle;
\end{scope}

\begin{scope}[shift=(r3Coord)]
  \draw[fill] circle (0.1);
\end{scope}

\begin{scope}[shift=(rLCoord), rotate=15]
  \draw[fill]
    plot[domain=0:360, samples=360, variable=\t]
      ({0.2*abs(cos(\t))*cos(\t)},
       {0.2*abs(cos(\t))*sin(\t)})
    -- cycle;
\end{scope}

\fill (r1Coord) node[above right=0.8mm]       {$\mathbf{r}_1$};
\fill (r2Coord) node[below left=2mm]  {$\mathbf{r}_3$};
\fill (r3Coord) node[above left=1mm]  {$\mathbf{r}_2$};
\fill (rLCoord) node[below right=1mm] {$\mathbf{r}_M$};

\node[font=\small, above left=-3pt]  at (arc1End) {$\phi_1$};
\node[font=\small, below right=-3pt] at (arc2End) {$\phi_3$};
\node[font=\small, below left=-3pt]  at (arc3End) {$\phi_2$};
\node[font=\small, above right=-3pt] at (arcMEnd) {$\phi_M$};

\end{tikzpicture}
    \caption{Illustration of an arbitrary array geometry. Each element $m$ is characterized by its own position $\mathbf{r}_m$ and directivity pattern $T_m(\omega, \theta)$.}
    \label{fig:plane}
\end{figure}

An illustrative example of the class of considered arrays is shown in Fig.~\ref{fig:plane}, with each array element arbitrarily positioned in the plane and characterized by a specific directivity $T_m(\omega, \theta)$. 
From now on, we omit the explicit dependence of $q_m$ on the angular frequency $\omega$. This simplification has no impact on the subsequent derivation, as the formulation is developed frequency by frequency and therefore remains valid when $q_m$ varies with $\omega$.

In this scenario, we can rewrite \eqref{eq:captured_signal} as
\begin{equation}\label{eq:captured_signal_directivity}
    Y_m(\omega) = T_m(\theta)D_m(\omega, \theta_{\mathrm{s}})X(\omega) + V_m(\omega),
\end{equation}
and \eqref{eq:captured_signal_matrix} as
\begin{equation}\label{eq:captured_signal_directivity_matrix}
    \mathbf{y}(\omega) = \mathbf{T}(\theta)\mathbf{d}(\omega, \theta_{\mathrm{s}})X(\omega) + \mathbf{v}(\omega),
\end{equation}
in which we have $\mathbf{T}(\theta) = \operatorname{diag}\left(T_1(\theta), \dots, T_M(\theta)\right)$.
In order to perform spatial filtering as in \eqref{eq:filtering}, we now need to account for the inherent directionality of first-order array elements, while designing the beamforming filter $\mathbf{h}(\omega)$.

\subsection{Metrics}\label{sec:metrics_new}
With the introduction of directional dependence in each array element, the performance metrics outlined in Sec.~\ref{sec:metrics} must be revised to reflect this change. In particular, the beampattern must now account for the individual directivity patterns of the array elements. It is therefore redefined as
\begin{equation}\label{eq:beampattern_directivity}
\begin{aligned}
    \mathcal{B}[\mathbf{h}(\omega),\theta] = \mathbf{h}^H(\omega)\mathbf{T}(\theta)\mathbf{d}(\omega, \theta).
\end{aligned}
\end{equation}
It is worth noting that if $\mathbf{T}(\theta) = \mathbf{I}$, i.e., if all elements are omnidirectional, \eqref{eq:beampattern_directivity} reduces to \eqref{eq:beampattern}, and the two definitions become equivalent.

Since both the white noise gain, as defined in \eqref{eq:wng}, and the directivity factor, as defined in \eqref{eq:df}, are expressed in terms of the beampattern, their formal definitions remain unchanged.
However, their numerical values will now reflect the influence of the directional patterns embedded in $\mathbf{T}(\theta)$.

\begin{figure*}[!t]
\centering
\subfloat[]{
\newcommand\AddMicFromCSV[3]{%
  \addplot[solid, data cs=cart] 
    table[
      col sep=space,
      x expr={ #2*cos(#1) + \thisrow{mag}*cos(\thisrow{angle}) },
      y expr={ #2*sin(#1) + \thisrow{mag}*sin(\thisrow{angle}) }
    ] {#3};
  \addplot[only marks, mark=*, mark size=1pt, data cs=cart]
    coordinates {({#2*cos(#1)},{#2*sin(#1)})};
}

\begin{tikzpicture}
\begin{polaraxis}[
    width=0.3\linewidth,
    axis line style={draw=none},
    grid=both,
    axis equal image,
    xtick={0,30,...,330},              
    xticklabels={\SI{0}{\degree},\SI{30}{\degree},\SI{60}{\degree},\SI{90}{\degree},\SI{120}{\degree},\SI{150}{\degree},\SI{180}{\degree},\SI{210}{\degree},\SI{240}{\degree},\SI{270}{\degree},\SI{300}{\degree},\SI{330}{\degree}},
    scaled y ticks=false,
    ymin=0, ymax=0.025/0.004,
    ytick={0.01/0.004,0.02/0.004,0.025/0.004},
    yticklabels=\empty,
    ticklabel style={font=\small},
    legend style={
      at={(0.5,1.15)}, anchor=south,
      legend columns=2,
      /tikz/every even column/.append style={column sep=1em}
    },
]

\addlegendimage{solid, black}
\addlegendentry{$T_m(\theta)$};

\addlegendimage{only marks, mark=*, mark size=1pt, black}
\addlegendentry{$\mathbf{r}_m$};

\AddMicFromCSV{-167.7108}{0.0145990825025377/0.004}{images/geometry/mic01.csv}
\AddMicFromCSV{-103.1169}{0.0142765967306843/0.004}{images/geometry/mic02.csv}
\AddMicFromCSV{55.7637}{0.018802969804683/0.004}{images/geometry/mic03.csv}
\AddMicFromCSV{83.0915}{0.00119229311968417/0.004}{images/geometry/mic04.csv}
\AddMicFromCSV{-36.2055}{0.0156482845788683/0.004}{images/geometry/mic05.csv}
\AddMicFromCSV{146.3960}{0.0112315231281142/0.004}{images/geometry/mic06.csv}
\AddMicFromCSV{90.8070}{0.00984468611877353/0.004}{images/geometry/mic07.csv}
\AddMicFromCSV{-7.2106}{0.0181948560918001/0.004}{images/geometry/mic08.csv}
\AddMicFromCSV{-0.7369}{0.00939084511021848/0.004}{images/geometry/mic09.csv}

\node [font=\small, anchor=west, xshift=-17pt] at (axis cs:60,0.02/0.004) {\SI{0.02}{\meter}};

\end{polaraxis}
\end{tikzpicture}%
\label{fig:geometry}}
\hfil
\subfloat[]{\input{images/bp_freq_1/img}%
\label{fig:bp_freq_1}}
\hfil
\subfloat[]{\input{images/bp_freq_2/img}%
\label{fig:bp_freq_2}}
\caption{Setup and results of the experiment in Sec.~\ref{sec:performance_fixed_geometry}. (a) Arbitrary planar array geometry with $M = 9$ microphones and a minimum inter-element spacing of \SI{8}{\milli\meter}. The positions of the array elements are indicated by black dots, with the corresponding first-order polar patterns shown at each element location. (b) First-order supercardioid and (c) fourth-order hypercardioid beampatterns (solid blue line), synthesized using the array configuration in (a), with target steering directions $\theta_{\mathrm{s}} = \frac{\pi}{3}$. In both cases, the overlaid target beampatterns (dashed red line) confirm that the proposed method closely approximates the desired spatial response across the evaluated frequency range.}
\label{fig:bp_frequency}
\end{figure*}

\subsection{Filter Design}
In the case of an array consisting of $M$ microphones with arbitrary planar positions and first-order directional responses, the beamformer design problem in \eqref{eq:renderedbp-targetbp} can be reformulated as
\begin{equation}\label{eq:renderedbp-targetbp-directional}
    \begin{aligned}
        \sum_{m=1}^M H_m^*(\omega)T_m(\theta)e^{jkr_m\cos(\theta - \phi_m)}
        = \sum_{n=-N}^N b_n e^{jn(\theta - \theta_{\mathrm{s}})}.
    \end{aligned}
\end{equation}
However, the dependence on $\theta$ of the  directivity term $T_m(\theta)$ does not allow to exploit the Jacobi-Anger expansion described in \eqref{eq:ja_expansion} to solve the modal-matching problem.
Similarly to \cite{wang2024design}, in which, however, the authors limit themselves to specific planar geometries (namely circular concentric arrays in which first-order elements are steered in the outward radial direction), we propose a more general series expansion for the array-element transfer function.
To this end, since the Jacobi-Anger expansion is a special case of the Fourier expansion \cite{williams1999fourier}, in order to solve the problem in \eqref{eq:renderedbp-targetbp-directional}, we expand $T_m(\theta)e^{jkr_m \cos(\theta-\phi_m)}$ by means of the circular harmonic decomposition \cite{teutsch2007modal} as
\begin{equation}\label{eq:fourier_expansion}
T_m(\theta)e^{jkr_m\cos(\theta-\phi_m)} = \sum_{n = -\infty}^{+\infty} \xi_{n,m} e^{jn(\theta-\theta_m)},
\end{equation}
where $\xi_{n,m}$ denotes the complex $n$th-order Fourier coefficient associated with the directional response of the $m$th array element. The problem thus reduces to determining the harmonic coefficients needed for the filter design, which can be computed as \cite{teutsch2007modal}
\begin{equation}\label{eq:fourier_coefficients}
    \begin{aligned}
    & \xi_{n,m} = \frac{1}{2\pi}\int_{-\pi}^{\pi}T_m(\theta)e^{jkr_m\cos(\theta-\phi_m)}e^{-jn(\theta-\theta_m)}\,d\theta.
\end{aligned}
\end{equation}

\begin{figure*}[b]
\smallskip
\hrule
\vspace{5mm}
\begin{dmath}
\xi_{n,m} = (1-q_m)\,
j^nJ_n(kr_m) e^{jn(\theta_m-\phi_m)}
+ \, q_m \,
j^{n+1} \left(\frac{ J_{n+1}(kr_m)e^{j(n+1)(\theta_m-\phi_m)} - J_{n-1}(kr_m)e^{j(n-1)(\theta_m-\phi_m)}}{2} \right)
\label{eq:fourier_coefficients_derived}
\end{dmath}
\end{figure*}

As demonstrated in the Appendix, from \eqref{eq:fourier_coefficients} we can derive \eqref{eq:fourier_coefficients_derived}, which is the formula for the harmonic coefficients.
In particular, in \eqref{eq:fourier_coefficients_derived} we observe the presence of two terms. The first one is weighted by $(1-q_m)$ and is related to the approximation of the omnidirectional component in \eqref{eq:transfer_function}. The second one, weighted by $q_m$, is related to the approximation of the steered dipole component.

At this point, by truncating \eqref{eq:fourier_expansion} to expansion order $N$, \eqref{eq:renderedbp-targetbp-directional} can be rewritten as
\begin{equation}\label{eq:renderedbp-targetbp-directional-fourier}
    \begin{aligned}
        \sum_{m=1}^M H_m^*(\omega)\sum_{n = -N}^{N} \xi_{n,m} e^{jn(\theta-\theta_m)}
        = \sum_{n=-N}^N b_n e^{jn(\theta - \theta_{\mathrm{s}})}.
    \end{aligned}
\end{equation}
Following the approach presented in Sec.~\ref{sec:filter_design}, it is now possible to drop the dependence on $\theta$ from both sides of \eqref{eq:renderedbp-targetbp-directional-fourier}, thus enabling us to exploit a modal-matching approach as
\begin{equation}\label{eq:modal-matching-fourier}
\sum_{m=1}^M H_m^*(\omega)\xi_{n,m} e^{-jn\theta_m} = b_n e^{-jn\theta_{\mathrm{s}}}.
\end{equation}
For each $n \in \{ -N, \dots, N\}$, \eqref{eq:modal-matching-fourier} can be expressed in matrix notation as
\begin{equation}\label{eq:modal_matching_fo}
    \boldsymbol{\Xi}(\omega)\mathbf{h}^*(\omega) = \mathbf{\Upsilon}(\theta_{\mathrm{s}})\mathbf{b},
\end{equation}
where $\boldsymbol{\Xi}(\omega)$ is a $(2N+1) \times M$ matrix defined as
\begin{equation}\label{eq:xi_matrix}
\begin{aligned}
\boldsymbol{\Xi}(\omega) = \begin{bmatrix}
        \xi_{-N,1}e^{jN\theta_1} & \dots & \xi_{-N,M}e^{jN\theta_M} \\
        \vdots & \ddots & \vdots \\
        \xi_{N,1}e^{-jN\theta_1} & \dots & \xi_{N,M}e^{-jN\theta_M} \\
    \end{bmatrix}.
\end{aligned}
\end{equation}
After conjugating both sides of \eqref{eq:modal_matching_fo} and obtaining $\boldsymbol{\Xi}^*(\omega)\mathbf{h}(\omega) = \mathbf{\Upsilon}^*(\theta_{\mathrm{s}})\mathbf{b}$, assuming $M \geq 2N+1$ and $\boldsymbol{\Xi}(\omega)$ has full row rank \cite{borra2020efficient}, the beamforming filter $\mathbf{h}(\omega)$ can be obtained as the minimum-norm solution
\begin{equation}\label{eq:minimum_norm_solution}
    \mathbf{h}(\omega) = [\boldsymbol{\Xi}^*(\omega)]^\dagger\mathbf{\Upsilon}^*(\theta_{\mathrm{s}})\mathbf{b}.
\end{equation}


It can be verified that \eqref{eq:xi_matrix} reduces to \eqref{eq:psi_matrix} when the array elements are omnidirectional (i.e., $q_m = 0$ for each microphone $m$). This confirms that the proposed method generalizes the FIB-LSE approach in \cite{huang2018designArbitrary} by incorporating the first-order directivity components of the array elements into the modal representation via \eqref{eq:fourier_coefficients_derived}.
Moreover, the proposed formulation also generalizes \cite{wang2024design}.
In particular, it can be verified that when each sensor's steering direction coincides with its angular position (i.e., $\theta_m = \phi_m$) and the array geometry is restricted to uniform concentric circular rings, \eqref{eq:fourier_coefficients_derived} reduces to the circular harmonic coefficients derived by Wang et al.\ \cite{wang2024design}.
\begin{figure*}[!t]
\centering
\subfloat[]{\begin{tikzpicture}
  \begin{polaraxis}[
    width=0.32\linewidth,
    height=0.32\linewidth,
    domain=0:360,
    samples=361,
    axis equal image,
    xtick={0,30,...,330},
    xticklabels={\SI{0}{\degree},\SI{30}{\degree},\SI{60}{\degree},\SI{90}{\degree},\SI{120}{\degree},\SI{150}{\degree},\SI{180}{\degree},\SI{210}{\degree},\SI{240}{\degree},\SI{270}{\degree},\SI{300}{\degree},\SI{330}{\degree},\SI{360}{\degree}},
    yticklabel style={rotate=45,anchor=west},
    ymin=-50,                           
    ymax=12,                           
    y coord trafo/.code=\pgfmathparse{#1 + 50.00001},
    y coord inv trafo/.code=\pgfmathparse{#1 - 50.00001},
    ytick={-40,-20,0,12},
    yticklabels=\empty,
    minor y tick num=1,                
    grid=both,
    axis line style={draw=none},
    tick label style={font=\small},
    legend style={
      at={(0.5,1.15)}, 
      anchor=south, 
      legend columns=2,   
      /tikz/every even column/.append style={column sep=1em} 
    },
  ]
    \addplot[thick, blue] table [x=angle, y=rendered] {images/bp_fo_m3/data.csv};
    \addlegendentry{$\mathcal{B}[\mathbf{h}(\omega),\theta]$}
    \addplot[thick, densely dashed, red] table [x=angle, y=target] {images/bp_fo_m3/data.csv};
    \addlegendentry{$\bar{\mathcal{B}}(\mathbf{b}, \theta, \theta_{\protect\mathrm{s}})$}
    \addplot[name path=ci_lower, draw=none] table [x=angle, y=lower_ci] {images/bp_fo_m3/data.csv};
    \addplot[name path=ci_upper, draw=none] table [x=angle, y=upper_ci] {images/bp_fo_m3/data.csv};
    \addplot[gray, fill opacity=0.5] fill between[of=ci_lower and ci_upper,];
    \node [font=\small, anchor=west, xshift=-17pt]at (axis cs:60,-40) {\SI{-40}{\decibel}};
    \node [font=\small, anchor=west, xshift=-17pt]at (axis cs:60,-20) {\SI{-20}{\decibel}};
    \node [font=\small, anchor=west, xshift=-6pt]at (axis cs:60,0) {\SI{0}{\decibel}};
\end{polaraxis}
\end{tikzpicture}%
\label{fig:bp_fo_m3}}
\hfil
\subfloat[]{\begin{tikzpicture}
  \begin{polaraxis}[
    width=0.32\linewidth,
    height=0.32\linewidth,
    domain=0:360,
    samples=361,
    axis equal image,
    xtick={0,30,...,330},
    xticklabels={\SI{0}{\degree},\SI{30}{\degree},\SI{60}{\degree},\SI{90}{\degree},\SI{120}{\degree},\SI{150}{\degree},\SI{180}{\degree},\SI{210}{\degree},\SI{240}{\degree},\SI{270}{\degree},\SI{300}{\degree},\SI{330}{\degree},\SI{360}{\degree}},
    yticklabel style={rotate=45,anchor=west},
    ymin=-50,                           
    ymax=12,                           
    y coord trafo/.code=\pgfmathparse{#1 + 50.00001},
    y coord inv trafo/.code=\pgfmathparse{#1 - 50.00001},
    ytick={-40,-20,0,12},
    yticklabels=\empty,
    minor y tick num=1,                
    grid=both,
    axis line style={draw=none},
    tick label style={font=\small},
    legend style={
      at={(0.5,1.15)},    
      anchor=south,       
      legend columns=2,   
      /tikz/every even column/.append style={column sep=1em} 
    },
  ]
    \addplot[thick, blue] table [x=angle, y=rendered] {images/bp_so_m5/data.csv};
    \addlegendentry{$\mathcal{B}[\mathbf{h}(\omega),\theta]$}
    \addplot[thick, densely dashed, red] table [x=angle, y=target] {images/bp_so_m5/data.csv};
    \addlegendentry{$\bar{\mathcal{B}}(\mathbf{b}, \theta, \theta_{\protect\mathrm{s}})$}
    \addplot[name path=ci_lower, draw=none] table [x=angle, y=lower_ci] {images/bp_so_m5/data.csv};
    \addplot[name path=ci_upper, draw=none] table [x=angle, y=upper_ci] {images/bp_so_m5/data.csv};
    \addplot[gray, fill opacity=0.5] fill between[of=ci_lower and ci_upper,];
    \node [font=\small, anchor=west, xshift=-17pt]at (axis cs:60,-40) {\SI{-40}{\decibel}};
    \node [font=\small, anchor=west, xshift=-17pt]at (axis cs:60,-20) {\SI{-20}{\decibel}};
    \node [font=\small, anchor=west, xshift=-6pt]at (axis cs:60,0) {\SI{0}{\decibel}};
\end{polaraxis}
\end{tikzpicture}%
\label{fig:bp_so_m5}}
\hfil
\subfloat[]{\begin{tikzpicture}
  \begin{polaraxis}[
    width=0.32\linewidth,
    height=0.32\linewidth,
    domain=0:360,
    samples=361,
    axis equal image,
    xtick={0,30,...,330},
    xticklabels={\SI{0}{\degree},\SI{30}{\degree},\SI{60}{\degree},\SI{90}{\degree},\SI{120}{\degree},\SI{150}{\degree},\SI{180}{\degree},\SI{210}{\degree},\SI{240}{\degree},\SI{270}{\degree},\SI{300}{\degree},\SI{330}{\degree},\SI{360}{\degree}},
    yticklabel style={rotate=45,anchor=west},
    ymin=-50,                           
    ymax=12,                           
    y coord trafo/.code=\pgfmathparse{#1 + 50.00001},
    y coord inv trafo/.code=\pgfmathparse{#1 - 50.00001},
    ytick={-40,-20,0,12},
    yticklabels=\empty,
    minor y tick num=1,                
    grid=both,
    axis line style={draw=none},
    tick label style={font=\small},
    legend style={
      at={(0.5,1.15)},    
      anchor=south,       
      legend columns=2,   
      /tikz/every even column/.append style={column sep=1em} 
    },
  ]
    \addplot[thick, blue] table [x=angle, y=rendered] {images/bp_to_m7/data.csv};
    \addlegendentry{$\mathcal{B}[\mathbf{h}(\omega),\theta]$}
    \addplot[thick, densely dashed, red] table [x=angle, y=target] {images/bp_to_m7/data.csv};
    \addlegendentry{$\bar{\mathcal{B}}(\mathbf{b}, \theta, \theta_{\protect\mathrm{s}})$}
    \addplot[name path=ci_lower, draw=none] table [x=angle, y=lower_ci] {images/bp_to_m7/data.csv};
    \addplot[name path=ci_upper, draw=none] table [x=angle, y=upper_ci] {images/bp_to_m7/data.csv};
    \addplot[gray, fill opacity=0.5] fill between[of=ci_lower and ci_upper,];
    \node [font=\small, anchor=west, xshift=-17pt]at (axis cs:60,-40) {\SI{-40}{\decibel}};
    \node [font=\small, anchor=west, xshift=-17pt]at (axis cs:60,-20) {\SI{-20}{\decibel}};
    \node [font=\small, anchor=west, xshift=-6pt]at (axis cs:60,0) {\SI{0}{\decibel}};
\end{polaraxis}
\end{tikzpicture}%
\label{fig:bp_to_m7}}\\
\subfloat[]{
\definecolor{mplblue}{RGB}{31,119,180}
\definecolor{mplorange}{RGB}{255,127,14}
\definecolor{mplyellow}{RGB}{255,191,0}

\pgfmathsetmacro{\bandfactor}{1.0}

\begin{tikzpicture}
\begin{axis}[
  width=0.47\linewidth, height=0.3\linewidth,
  xmin=0.05, xmax=4, ymin=-80, ymax=20,
  xtick distance=0.5,
  grid=both, 
  xlabel={Frequency (\si{\kilo\hertz})},
  ylabel={$\mathcal{W}$ (\si{\decibel})},
  legend style={at={(0.47,0.07)}, anchor=south west},
  legend cell align=left,
  tick label style={/pgf/number format/fixed},
]

\addlegendimage{line legend, thick, draw=mplblue}
\addlegendentry{$N=1,\ M=3$}

\addlegendimage{line legend, thick, dashed, draw=mplorange}
\addlegendentry{$N=2,\ M=5$}

\addlegendimage{line legend, thick, dotted, draw=mplyellow}
\addlegendentry{$N=3,\ M=7$}

\addplot[name path=hiA, draw=none]
  table[col sep=space,
        x expr=\thisrow{freq}/1000,
        y expr=\thisrow{mean_wng} + \bandfactor*\thisrow{std_wng}]
        {images/wng_m357/data_m3.csv};
\addplot[name path=loA, draw=none]
  table[col sep=space,
        x expr=\thisrow{freq}/1000,
        y expr=\thisrow{mean_wng} - \bandfactor*\thisrow{std_wng}]
        {images/wng_m357/data_m3.csv};
\addplot[forget plot, fill=mplblue, fill opacity=0.2]
  fill between[of=hiA and loA];
\addplot[thick, mplblue]
  table[col sep=space, x expr=\thisrow{freq}/1000, y=mean_wng]
        {images/wng_m357/data_m3.csv};

\addplot[name path=hiB, draw=none]
  table[col sep=space,
        x expr=\thisrow{freq}/1000,
        y expr=\thisrow{mean_wng} + \bandfactor*\thisrow{std_wng}]
        {images/wng_m357/data_m5.csv};
\addplot[name path=loB, draw=none]
  table[col sep=space,
        x expr=\thisrow{freq}/1000,
        y expr=\thisrow{mean_wng} - \bandfactor*\thisrow{std_wng}]
        {images/wng_m357/data_m5.csv};
\addplot[forget plot, fill=mplorange, fill opacity=0.2]
  fill between[of=hiB and loB];
\addplot[thick, draw=mplorange, dashed]
  table[col sep=space, x expr=\thisrow{freq}/1000, y=mean_wng]
        {images/wng_m357/data_m5.csv};

\addplot[name path=hiC, draw=none]
  table[col sep=space,
        x expr=\thisrow{freq}/1000,
        y expr=\thisrow{mean_wng} + \bandfactor*\thisrow{std_wng}]
        {images/wng_m357/data_m7.csv};
\addplot[name path=loC, draw=none]
  table[col sep=space,
        x expr=\thisrow{freq}/1000,
        y expr=\thisrow{mean_wng} - \bandfactor*\thisrow{std_wng}]
        {images/wng_m357/data_m7.csv};
\addplot[forget plot, fill=mplyellow, fill opacity=0.2]
  fill between[of=hiC and loC];
\addplot[thick, draw=mplyellow, dotted]
  table[col sep=space, x expr=\thisrow{freq}/1000, y=mean_wng]
        {images/wng_m357/data_m7.csv};

\end{axis}
\end{tikzpicture}%
\label{fig:wng_m357}}
\hfil
\subfloat[]{
\definecolor{mplblue}{RGB}{31,119,180}
\definecolor{mplorange}{RGB}{255,127,14}
\definecolor{mplyellow}{RGB}{255,191,0}

\pgfmathsetmacro{\bandfactor}{1.0}

\begin{tikzpicture}
\begin{axis}[
  width=0.47\linewidth, height=0.3\linewidth,
  xmin=0.05, xmax=4, ymin=0, ymax=20,
  xtick distance=0.5,
  grid=both, 
  xlabel={Frequency (\si{\kilo\hertz})},
  ylabel={$\mathcal{D}$ (\si{\decibel})},
  legend style={at={(0.47,0.07)}, anchor=south west},
  legend cell align=left,
  tick label style={/pgf/number format/fixed},
]

\addlegendimage{line legend, thick, draw=mplblue}
\addlegendentry{$N=1,\ M=3$}

\addlegendimage{line legend, thick, dashed, draw=mplorange}
\addlegendentry{$N=2,\ M=5$}

\addlegendimage{line legend, thick, dotted, draw=mplyellow}
\addlegendentry{$N=3,\ M=7$}

\addplot[name path=hiA, draw=none]
  table[col sep=space,
        x expr=\thisrow{freq}/1000,
        y expr=\thisrow{mean_df} + \bandfactor*\thisrow{std_df}]
        {images/df_m357/data_m3.csv};
\addplot[name path=loA, draw=none]
  table[col sep=space,
        x expr=\thisrow{freq}/1000,
        y expr=\thisrow{mean_df} - \bandfactor*\thisrow{std_df}]
        {images/df_m357/data_m3.csv};
\addplot[forget plot, fill=mplblue, fill opacity=0.2]
  fill between[of=hiA and loA];
\addplot[thick, mplblue]
  table[col sep=space, x expr=\thisrow{freq}/1000, y=mean_df]
        {images/df_m357/data_m3.csv};

\addplot[name path=hiB, draw=none]
  table[col sep=space,
        x expr=\thisrow{freq}/1000,
        y expr=\thisrow{mean_df} + \bandfactor*\thisrow{std_df}]
        {images/df_m357/data_m5.csv};
\addplot[name path=loB, draw=none]
  table[col sep=space,
        x expr=\thisrow{freq}/1000,
        y expr=\thisrow{mean_df} - \bandfactor*\thisrow{std_df}]
        {images/df_m357/data_m5.csv};
\addplot[forget plot, fill=mplorange, fill opacity=0.2]
  fill between[of=hiB and loB];
\addplot[thick, draw=mplorange, dashed]
  table[col sep=space, x expr=\thisrow{freq}/1000, y=mean_df]
        {images/df_m357/data_m5.csv};

\addplot[name path=hiC, draw=none]
  table[col sep=space,
        x expr=\thisrow{freq}/1000,
        y expr=\thisrow{mean_df} + \bandfactor*\thisrow{std_df}]
        {images/df_m357/data_m7.csv};
\addplot[name path=loC, draw=none]
  table[col sep=space,
        x expr=\thisrow{freq}/1000,
        y expr=\thisrow{mean_df} - \bandfactor*\thisrow{std_df}]
        {images/df_m357/data_m7.csv};
\addplot[forget plot, fill=mplyellow, fill opacity=0.2]
  fill between[of=hiC and loC];
\addplot[thick, draw=mplyellow, dotted]
  table[col sep=space, x expr=\thisrow{freq}/1000, y=mean_df]
        {images/df_m357/data_m7.csv};

\end{axis}
\end{tikzpicture}%
\label{fig:df_m357}}
\caption{Results of the Monte Carlo experiment in Sec.~\ref{sec:robustness_geometry_variations}. The top row shows the average (solid blue line) and target (dashed red line) hypercardioid beampatterns ($\theta_{\mathrm{s}} = \frac{\pi}{3}$) at \SI{1}{\kilo\hertz} for orders (a)~$N=1$, (b)~$N=2$, and (c)~$N=3$, synthesized using arbitrary arrays of $M=3,5,7$ microphones, respectively. The bottom row reports the corresponding average (d) white noise gain and (e) directivity factor as functions of frequency for the same target beampatterns and array configurations. Shaded regions denote one standard deviation around the mean. The results demonstrate accurate beampattern approximation and reduced sensitivity to array geometry variations, with performance improving as the number of microphones increases.}
\label{fig:montecarlo_m357}
\end{figure*}

\begin{figure*}[!t]
\centering
\subfloat[]{\begin{tikzpicture}
  \begin{polaraxis}[
    width=0.32\linewidth,
    height=0.32\linewidth,
    domain=0:360,
    samples=361,
    axis equal image,
    xtick={0,30,...,330},
    xticklabels={\SI{0}{\degree},\SI{30}{\degree},\SI{60}{\degree},\SI{90}{\degree},\SI{120}{\degree},\SI{150}{\degree},\SI{180}{\degree},\SI{210}{\degree},\SI{240}{\degree},\SI{270}{\degree},\SI{300}{\degree},\SI{330}{\degree},\SI{360}{\degree}},
    yticklabel style={rotate=45,anchor=west},
    ymin=-50,                           
    ymax=12,                           
    y coord trafo/.code=\pgfmathparse{#1 + 50.00001},
    y coord inv trafo/.code=\pgfmathparse{#1 - 50.00001},
    ytick={-40,-20,0,12},
    yticklabels=\empty,
    minor y tick num=1,                
    grid=both,
    axis line style={draw=none},
    tick label style={font=\small},
    legend style={
      at={(0.5,1.15)},    
      anchor=south,       
      legend columns=2,   
      /tikz/every even column/.append style={column sep=1em} 
    },
  ]
    \addplot[thick, blue] table [x=angle, y=rendered] {images/bp_so_m5_v2/data.csv};
    \addlegendentry{$\mathcal{B}[\mathbf{h}(\omega),\theta]$}
    \addplot[thick, densely dashed, red] table [x=angle, y=target] {images/bp_so_m5_v2/data.csv};
    \addlegendentry{$\bar{\mathcal{B}}(\mathbf{b}, \theta, \theta_{\protect\mathrm{s}})$}
    \addplot[name path=ci_lower, draw=none] table [x=angle, y=lower_ci] {images/bp_so_m5_v2/data.csv};
    \addplot[name path=ci_upper, draw=none] table [x=angle, y=upper_ci] {images/bp_so_m5_v2/data.csv};
    \addplot[gray, fill opacity=0.5] fill between[of=ci_lower and ci_upper,];
    \node [font=\small, anchor=west, xshift=-17pt]at (axis cs:60,-40) {\SI{-40}{\decibel}};
    \node [font=\small, anchor=west, xshift=-17pt]at (axis cs:60,-20) {\SI{-20}{\decibel}};
    \node [font=\small, anchor=west, xshift=-6pt]at (axis cs:60,0) {\SI{0}{\decibel}};
\end{polaraxis}
\end{tikzpicture}%
\label{fig:bp_so_m5_v2}}
\hfil
\subfloat[]{\begin{tikzpicture}
  \begin{polaraxis}[
    width=0.32\linewidth,
    height=0.32\linewidth,
    domain=0:360,
    samples=361,
    axis equal image,
    xtick={0,30,...,330},
    xticklabels={\SI{0}{\degree},\SI{30}{\degree},\SI{60}{\degree},\SI{90}{\degree},\SI{120}{\degree},\SI{150}{\degree},\SI{180}{\degree},\SI{210}{\degree},\SI{240}{\degree},\SI{270}{\degree},\SI{300}{\degree},\SI{330}{\degree},\SI{360}{\degree}},
    yticklabel style={rotate=45,anchor=west},
    ymin=-50,                           
    ymax=12,                           
    y coord trafo/.code=\pgfmathparse{#1 + 50.00001},
    y coord inv trafo/.code=\pgfmathparse{#1 - 50.00001},
    ytick={-40,-20,0,12},
    yticklabels=\empty,
    minor y tick num=1,                
    grid=both,
    axis line style={draw=none},
    tick label style={font=\small},
    legend style={
      at={(0.5,1.15)},    
      anchor=south,       
      legend columns=2,   
      /tikz/every even column/.append style={column sep=1em} 
    },
  ]
    \addplot[thick, blue] table [x=angle, y=rendered] {images/bp_so_m9_v2/data.csv};
    \addlegendentry{$\mathcal{B}[\mathbf{h}(\omega),\theta]$}
    \addplot[thick, densely dashed, red] table [x=angle, y=target] {images/bp_so_m9_v2/data.csv};
    \addlegendentry{$\bar{\mathcal{B}}(\mathbf{b}, \theta, \theta_{\protect\mathrm{s}})$}
    \addplot[name path=ci_lower, draw=none] table [x=angle, y=lower_ci] {images/bp_so_m9_v2/data.csv};
    \addplot[name path=ci_upper, draw=none] table [x=angle, y=upper_ci] {images/bp_so_m9_v2/data.csv};
    \addplot[gray, fill opacity=0.5] fill between[of=ci_lower and ci_upper,];
    \node [font=\small, anchor=west, xshift=-17pt]at (axis cs:60,-40) {\SI{-40}{\decibel}};
    \node [font=\small, anchor=west, xshift=-17pt]at (axis cs:60,-20) {\SI{-20}{\decibel}};
    \node [font=\small, anchor=west, xshift=-6pt]at (axis cs:60,0) {\SI{0}{\decibel}};
\end{polaraxis}
\end{tikzpicture}%
\label{fig:bp_so_m9_v2}}
\hfil
\subfloat[]{\begin{tikzpicture}
  \begin{polaraxis}[
    width=0.32\linewidth,
    height=0.32\linewidth,
    domain=0:360,
    samples=361,
    axis equal image,
    xtick={0,30,...,330},
    xticklabels={\SI{0}{\degree},\SI{30}{\degree},\SI{60}{\degree},\SI{90}{\degree},\SI{120}{\degree},\SI{150}{\degree},\SI{180}{\degree},\SI{210}{\degree},\SI{240}{\degree},\SI{270}{\degree},\SI{300}{\degree},\SI{330}{\degree},\SI{360}{\degree}},
    yticklabel style={rotate=45,anchor=west},
    ymin=-50,                           
    ymax=12,                           
    y coord trafo/.code=\pgfmathparse{#1 + 50.00001},
    y coord inv trafo/.code=\pgfmathparse{#1 - 50.00001},
    ytick={-40,-20,0,12},
    yticklabels=\empty,
    minor y tick num=1,                
    grid=both,
    axis line style={draw=none},
    tick label style={font=\small},
    legend style={
      at={(0.5,1.15)},    
      anchor=south,       
      legend columns=2,   
      /tikz/every even column/.append style={column sep=1em} 
    },
  ]
    \addplot[thick, blue] table [x=angle, y=rendered] {images/bp_so_m13/data.csv};
    \addlegendentry{$\mathcal{B}[\mathbf{h}(\omega),\theta]$}
    \addplot[thick, densely dashed, red] table [x=angle, y=target] {images/bp_so_m13/data.csv};
    \addlegendentry{$\bar{\mathcal{B}}(\mathbf{b}, \theta, \theta_{\protect\mathrm{s}})$}
    \addplot[name path=ci_lower, draw=none] table [x=angle, y=lower_ci] {images/bp_so_m13/data.csv};
    \addplot[name path=ci_upper, draw=none] table [x=angle, y=upper_ci] {images/bp_so_m13/data.csv};
    \addplot[gray, fill opacity=0.5] fill between[of=ci_lower and ci_upper,];
    \node [font=\small, anchor=west, xshift=-17pt]at (axis cs:60,-40) {\SI{-40}{\decibel}};
    \node [font=\small, anchor=west, xshift=-17pt]at (axis cs:60,-20) {\SI{-20}{\decibel}};
    \node [font=\small, anchor=west, xshift=-6pt]at (axis cs:60,0) {\SI{0}{\decibel}};
\end{polaraxis}
\end{tikzpicture}%
\label{fig:bp_so_m13}}\\
\subfloat[]{
\definecolor{mplblue}{RGB}{31,119,180}
\definecolor{mplorange}{RGB}{255,127,14}
\definecolor{mplyellow}{RGB}{255,191,0}

\pgfmathsetmacro{\bandfactor}{1.0}

\begin{tikzpicture}
\begin{axis}[
  width=0.47\linewidth, height=0.3\linewidth,
  xmin=0.05, xmax=4, ymin=-80, ymax=20,
  xtick distance=0.5,
  grid=both, 
  xlabel={Frequency (\si{\kilo\hertz})},
  ylabel={$\mathcal{W}$ (\si{\decibel})},
  legend style={at={(0.47,0.07)}, anchor=south west},
  legend cell align=left,
  tick label style={/pgf/number format/fixed},
]

\addlegendimage{line legend, thick, draw=mplblue}
\addlegendentry{$N=2,\ M=5$}

\addlegendimage{line legend, thick, dashed, draw=mplorange}
\addlegendentry{$N=2,\ M=9$}

\addlegendimage{line legend, thick, dotted, draw=mplyellow}
\addlegendentry{$N=2,\ M=13$}

\addplot[name path=hiA, draw=none]
  table[col sep=space,
        x expr=\thisrow{freq}/1000,
        y expr=\thisrow{mean_wng} + \bandfactor*\thisrow{std_wng}]
        {images/wng_m5913/data_m5_n2.csv};
\addplot[name path=loA, draw=none]
  table[col sep=space,
        x expr=\thisrow{freq}/1000,
        y expr=\thisrow{mean_wng} - \bandfactor*\thisrow{std_wng}]
        {images/wng_m5913/data_m5_n2.csv};
\addplot[forget plot, fill=mplblue, fill opacity=0.2]
  fill between[of=hiA and loA];
\addplot[thick, mplblue]
  table[col sep=space, x expr=\thisrow{freq}/1000, y=mean_wng]
        {images/wng_m5913/data_m5_n2.csv};

\addplot[name path=hiB, draw=none]
  table[col sep=space,
        x expr=\thisrow{freq}/1000,
        y expr=\thisrow{mean_wng} + \bandfactor*\thisrow{std_wng}]
        {images/wng_m5913/data_m9_n2.csv};
\addplot[name path=loB, draw=none]
  table[col sep=space,
        x expr=\thisrow{freq}/1000,
        y expr=\thisrow{mean_wng} - \bandfactor*\thisrow{std_wng}]
        {images/wng_m5913/data_m9_n2.csv};
\addplot[forget plot, fill=mplorange, fill opacity=0.2]
  fill between[of=hiB and loB];
\addplot[thick, draw=mplorange, dashed]
  table[col sep=space, x expr=\thisrow{freq}/1000, y=mean_wng]
        {images/wng_m5913/data_m9_n2.csv};

\addplot[name path=hiC, draw=none]
  table[col sep=space,
        x expr=\thisrow{freq}/1000,
        y expr=\thisrow{mean_wng} + \bandfactor*\thisrow{std_wng}]
        {images/wng_m5913/data_m13_n2.csv};
\addplot[name path=loC, draw=none]
  table[col sep=space,
        x expr=\thisrow{freq}/1000,
        y expr=\thisrow{mean_wng} - \bandfactor*\thisrow{std_wng}]
        {images/wng_m5913/data_m13_n2.csv};
\addplot[forget plot, fill=mplyellow, fill opacity=0.2]
  fill between[of=hiC and loC];
\addplot[thick, draw=mplyellow, dotted]
  table[col sep=space, x expr=\thisrow{freq}/1000, y=mean_wng]
        {images/wng_m5913/data_m13_n2.csv};

\end{axis}
\end{tikzpicture}%
\label{fig:wng_m5913}}
\hfil
\subfloat[]{
\definecolor{mplblue}{RGB}{31,119,180}
\definecolor{mplorange}{RGB}{255,127,14}
\definecolor{mplyellow}{RGB}{255,191,0}

\pgfmathsetmacro{\bandfactor}{1.0}

\begin{tikzpicture}
\begin{axis}[
  width=0.47\linewidth, height=0.3\linewidth,
  xmin=0.05, xmax=4, ymin=0, ymax=20,
  xtick distance=0.5,
  grid=both, 
  xlabel={Frequency (\si{\kilo\hertz})},
  ylabel={$\mathcal{D}$ (\si{\decibel})},
  legend style={at={(0.47,0.07)}, anchor=south west},
  legend cell align=left,
  tick label style={/pgf/number format/fixed},
]

\addlegendimage{line legend, thick, draw=mplblue}
\addlegendentry{$N=2,\ M=5$}

\addlegendimage{line legend, thick, dashed, draw=mplorange}
\addlegendentry{$N=2,\ M=9$}

\addlegendimage{line legend, thick, dotted, draw=mplyellow}
\addlegendentry{$N=2,\ M=13$}

\addplot[name path=hiA, draw=none]
  table[col sep=space,
        x expr=\thisrow{freq}/1000,
        y expr=\thisrow{mean_df} + \bandfactor*\thisrow{std_df}]
        {images/df_m5913/data_m5_n2.csv};
\addplot[name path=loA, draw=none]
  table[col sep=space,
        x expr=\thisrow{freq}/1000,
        y expr=\thisrow{mean_df} - \bandfactor*\thisrow{std_df}]
        {images/df_m5913/data_m5_n2.csv};
\addplot[forget plot, fill=mplblue, fill opacity=0.2]
  fill between[of=hiA and loA];
\addplot[thick, mplblue]
  table[col sep=space, x expr=\thisrow{freq}/1000, y=mean_df]
        {images/df_m5913/data_m5_n2.csv};

\addplot[name path=hiB, draw=none]
  table[col sep=space,
        x expr=\thisrow{freq}/1000,
        y expr=\thisrow{mean_df} + \bandfactor*\thisrow{std_df}]
        {images/df_m5913/data_m9_n2.csv};
\addplot[name path=loB, draw=none]
  table[col sep=space,
        x expr=\thisrow{freq}/1000,
        y expr=\thisrow{mean_df} - \bandfactor*\thisrow{std_df}]
        {images/df_m5913/data_m9_n2.csv};
\addplot[forget plot, fill=mplorange, fill opacity=0.2]
  fill between[of=hiB and loB];
\addplot[thick, draw=mplorange, dashed]
  table[col sep=space, x expr=\thisrow{freq}/1000, y=mean_df]
        {images/df_m5913/data_m9_n2.csv};

\addplot[name path=hiC, draw=none]
  table[col sep=space,
        x expr=\thisrow{freq}/1000,
        y expr=\thisrow{mean_df} + \bandfactor*\thisrow{std_df}]
        {images/df_m5913/data_m13_n2.csv};
\addplot[name path=loC, draw=none]
  table[col sep=space,
        x expr=\thisrow{freq}/1000,
        y expr=\thisrow{mean_df} - \bandfactor*\thisrow{std_df}]
        {images/df_m5913/data_m13_n2.csv};
\addplot[forget plot, fill=mplyellow, fill opacity=0.2]
  fill between[of=hiC and loC];
\addplot[thick, draw=mplyellow, dotted]
  table[col sep=space, x expr=\thisrow{freq}/1000, y=mean_df]
        {images/df_m5913/data_m13_n2.csv};

\end{axis}
\end{tikzpicture}%
\label{fig:df_m5913}}
\caption{Results of the Monte Carlo experiment in Sec.~\ref{sec:performance_increasing_elements}. The top row shows the average (solid blue line) and target (dashed red line) second-order supercardioid beampattern ($\theta_{\mathrm{s}} = \frac{\pi}{3}$) at \SI{1}{\kilo\hertz}, synthesized using arbitrary arrays of (a)~$M=5$, (b)~$M=9$, and (c)~$M=13$ microphones. The bottom row reports the corresponding average (d) white noise gain and (e) directivity factor as functions of frequency for the same target beampatterns and array configurations. Shaded regions denote one standard deviation around the mean. As the array size grows, beampattern approximation accuracy improves and white noise gain increases, whereas the directivity factor remains unchanged.}
\label{fig:montecarlo_m5913}
\end{figure*}

\begin{figure*}[!t]
\centering
\subfloat[]{\begin{tikzpicture}
  \begin{polaraxis}[
    width=0.32\linewidth,
    height=0.32\linewidth,
    domain=0:360,
    samples=361,
    axis equal image,
    xtick={0,30,...,330},
    xticklabels={\SI{0}{\degree},\SI{30}{\degree},\SI{60}{\degree},\SI{90}{\degree},\SI{120}{\degree},\SI{150}{\degree},\SI{180}{\degree},\SI{210}{\degree},\SI{240}{\degree},\SI{270}{\degree},\SI{300}{\degree},\SI{330}{\degree},\SI{360}{\degree}},
    yticklabel style={rotate=45,anchor=west},
    ymin=-50,                           
    ymax=12,                           
    y coord trafo/.code=\pgfmathparse{#1 + 50.00001},
    y coord inv trafo/.code=\pgfmathparse{#1 - 50.00001},
    ytick={-40,-20,0,12},
    yticklabels=\empty,
    minor y tick num=1,                
    grid=both,
    axis line style={draw=none},
    tick label style={font=\small},
    legend style={
      at={(0.5,1.15)},    
      anchor=south,       
      legend columns=2,   
      /tikz/every even column/.append style={column sep=1em} 
    },
  ]
    \addplot[thick, blue] table [x=angle, y=rendered] {images/bp_fo_m9/data.csv};
    \addlegendentry{$\mathcal{B}[\mathbf{h}(\omega),\theta]$}
    \addplot[thick, densely dashed, red] table [x=angle, y=target] {images/bp_fo_m9/data.csv};
    \addlegendentry{$\bar{\mathcal{B}}(\mathbf{b}, \theta, \theta_{\protect\mathrm{s}})$}
    \addplot[name path=ci_lower, draw=none] table [x=angle, y=lower_ci] {images/bp_fo_m9/data.csv};
    \addplot[name path=ci_upper, draw=none] table [x=angle, y=upper_ci] {images/bp_fo_m9/data.csv};
    \addplot[gray, fill opacity=0.5] fill between[of=ci_lower and ci_upper,];
    \node [font=\small, anchor=west, xshift=-17pt]at (axis cs:60,-40) {\SI{-40}{\decibel}};
    \node [font=\small, anchor=west, xshift=-17pt]at (axis cs:60,-20) {\SI{-20}{\decibel}};
    \node [font=\small, anchor=west, xshift=-6pt]at (axis cs:60,0) {\SI{0}{\decibel}};
\end{polaraxis}
\end{tikzpicture}%
\label{fig:bp_fo_m9}}
\hfil
\subfloat[]{\begin{tikzpicture}
  \begin{polaraxis}[
    width=0.32\linewidth,
    height=0.32\linewidth,
    domain=0:360,
    samples=361,
    axis equal image,
    xtick={0,30,...,330},
    xticklabels={\SI{0}{\degree},\SI{30}{\degree},\SI{60}{\degree},\SI{90}{\degree},\SI{120}{\degree},\SI{150}{\degree},\SI{180}{\degree},\SI{210}{\degree},\SI{240}{\degree},\SI{270}{\degree},\SI{300}{\degree},\SI{330}{\degree},\SI{360}{\degree}},
    yticklabel style={rotate=45,anchor=west},
    ymin=-50,                           
    ymax=12,                           
    y coord trafo/.code=\pgfmathparse{#1 + 50.00001},
    y coord inv trafo/.code=\pgfmathparse{#1 - 50.00001},
    ytick={-40,-20,0,12},
    yticklabels=\empty,
    minor y tick num=1,                
    grid=both,
    axis line style={draw=none},
    tick label style={font=\small},
    legend style={
      at={(0.5,1.15)},    
      anchor=south,       
      legend columns=2,   
      /tikz/every even column/.append style={column sep=1em} 
    },
  ]
    \addplot[thick, blue] table [x=angle, y=rendered] {images/bp_so_m9/data.csv};
    \addlegendentry{$\mathcal{B}[\mathbf{h}(\omega),\theta]$}
    \addplot[thick, densely dashed, red] table [x=angle, y=target] {images/bp_so_m9/data.csv};
    \addlegendentry{$\bar{\mathcal{B}}(\mathbf{b}, \theta, \theta_{\protect\mathrm{s}})$}
    \addplot[name path=ci_lower, draw=none] table [x=angle, y=lower_ci] {images/bp_so_m9/data.csv};
    \addplot[name path=ci_upper, draw=none] table [x=angle, y=upper_ci] {images/bp_so_m9/data.csv};
    \addplot[gray, fill opacity=0.5] fill between[of=ci_lower and ci_upper,];
    \node [font=\small, anchor=west, xshift=-17pt]at (axis cs:60,-40) {\SI{-40}{\decibel}};
    \node [font=\small, anchor=west, xshift=-17pt]at (axis cs:60,-20) {\SI{-20}{\decibel}};
    \node [font=\small, anchor=west, xshift=-6pt]at (axis cs:60,0) {\SI{0}{\decibel}};
\end{polaraxis}
\end{tikzpicture}%
\label{fig:bp_so_m9}}
\hfil
\subfloat[]{\begin{tikzpicture}
  \begin{polaraxis}[
    width=0.32\linewidth,
    height=0.32\linewidth,
    domain=0:360,
    samples=361,
    axis equal image,
    xtick={0,30,...,330},
    xticklabels={\SI{0}{\degree},\SI{30}{\degree},\SI{60}{\degree},\SI{90}{\degree},\SI{120}{\degree},\SI{150}{\degree},\SI{180}{\degree},\SI{210}{\degree},\SI{240}{\degree},\SI{270}{\degree},\SI{300}{\degree},\SI{330}{\degree},\SI{360}{\degree}},
    yticklabel style={rotate=45,anchor=west},
    ymin=-50,                           
    ymax=12,                           
    y coord trafo/.code=\pgfmathparse{#1 + 50.00001},
    y coord inv trafo/.code=\pgfmathparse{#1 - 50.00001},
    ytick={-40,-20,0,12},
    yticklabels=\empty,
    minor y tick num=1,                
    grid=both,
    axis line style={draw=none},
    tick label style={font=\small},
    legend style={
      at={(0.5,1.15)},    
      anchor=south,       
      legend columns=2,   
      /tikz/every even column/.append style={column sep=1em} 
    },
  ]
    \addplot[thick, blue] table [x=angle, y=rendered] {images/bp_to_m9/data.csv};
    \addlegendentry{$\mathcal{B}[\mathbf{h}(\omega),\theta]$}
    \addplot[thick, densely dashed, red] table [x=angle, y=target] {images/bp_to_m9/data.csv};
    \addlegendentry{$\bar{\mathcal{B}}(\mathbf{b}, \theta, \theta_{\protect\mathrm{s}})$}
    \addplot[name path=ci_lower, draw=none] table [x=angle, y=lower_ci] {images/bp_to_m9/data.csv};
    \addplot[name path=ci_upper, draw=none] table [x=angle, y=upper_ci] {images/bp_to_m9/data.csv};
    \addplot[gray, fill opacity=0.5] fill between[of=ci_lower and ci_upper,];
    \node [font=\small, anchor=west, xshift=-17pt]at (axis cs:60,-40) {\SI{-40}{\decibel}};
    \node [font=\small, anchor=west, xshift=-17pt]at (axis cs:60,-20) {\SI{-20}{\decibel}};
    \node [font=\small, anchor=west, xshift=-6pt]at (axis cs:60,0) {\SI{0}{\decibel}};
\end{polaraxis}
\end{tikzpicture}%
\label{fig:bp_to_m9}}\\
\subfloat[]{
\definecolor{mplblue}{RGB}{31,119,180}
\definecolor{mplorange}{RGB}{255,127,14}
\definecolor{mplyellow}{RGB}{255,191,0}

\pgfmathsetmacro{\bandfactor}{1.0}

\begin{tikzpicture}
\begin{axis}[
  width=0.47\linewidth, height=0.3\linewidth,
  xmin=0.05, xmax=4, ymin=-80, ymax=20,
  xtick distance=0.5,
  grid=both, 
  xlabel={Frequency (\si{\kilo\hertz})},
  ylabel={$\mathcal{W}$ (\si{\decibel})},
  legend style={at={(0.47,0.07)}, anchor=south west},
  legend cell align=left,
  tick label style={/pgf/number format/fixed},
]

\addlegendimage{line legend, thick, draw=mplblue}
\addlegendentry{$N=1,\ M=9$}

\addlegendimage{line legend, thick, dashed, draw=mplorange}
\addlegendentry{$N=2,\ M=9$}

\addlegendimage{line legend, thick, dotted, draw=mplyellow}
\addlegendentry{$N=3,\ M=9$}

\addplot[name path=hiA, draw=none]
  table[col sep=space,
        x expr=\thisrow{freq}/1000,
        y expr=\thisrow{mean_wng} + \bandfactor*\thisrow{std_wng}]
        {images/wng_m999/data_m9_n1.csv};
\addplot[name path=loA, draw=none]
  table[col sep=space,
        x expr=\thisrow{freq}/1000,
        y expr=\thisrow{mean_wng} - \bandfactor*\thisrow{std_wng}]
        {images/wng_m999/data_m9_n1.csv};
\addplot[forget plot, fill=mplblue, fill opacity=0.2]
  fill between[of=hiA and loA];
\addplot[thick, mplblue]
  table[col sep=space, x expr=\thisrow{freq}/1000, y=mean_wng]
        {images/wng_m999/data_m9_n1.csv};

\addplot[name path=hiB, draw=none]
  table[col sep=space,
        x expr=\thisrow{freq}/1000,
        y expr=\thisrow{mean_wng} + \bandfactor*\thisrow{std_wng}]
        {images/wng_m999/data_m9_n2.csv};
\addplot[name path=loB, draw=none]
  table[col sep=space,
        x expr=\thisrow{freq}/1000,
        y expr=\thisrow{mean_wng} - \bandfactor*\thisrow{std_wng}]
        {images/wng_m999/data_m9_n2.csv};
\addplot[forget plot, fill=mplorange, fill opacity=0.2]
  fill between[of=hiB and loB];
\addplot[thick, draw=mplorange, dashed]
  table[col sep=space, x expr=\thisrow{freq}/1000, y=mean_wng]
        {images/wng_m999/data_m9_n2.csv};

\addplot[name path=hiC, draw=none]
  table[col sep=space,
        x expr=\thisrow{freq}/1000,
        y expr=\thisrow{mean_wng} + \bandfactor*\thisrow{std_wng}]
        {images/wng_m999/data_m9_n3.csv};
\addplot[name path=loC, draw=none]
  table[col sep=space,
        x expr=\thisrow{freq}/1000,
        y expr=\thisrow{mean_wng} - \bandfactor*\thisrow{std_wng}]
        {images/wng_m999/data_m9_n3.csv};
\addplot[forget plot, fill=mplyellow, fill opacity=0.2]
  fill between[of=hiC and loC];
\addplot[thick, draw=mplyellow, dotted]
  table[col sep=space, x expr=\thisrow{freq}/1000, y=mean_wng]
        {images/wng_m999/data_m9_n3.csv};

\end{axis}
\end{tikzpicture}%
\label{fig:wng_m999}}
\hfil
\subfloat[]{
\definecolor{mplblue}{RGB}{31,119,180}
\definecolor{mplorange}{RGB}{255,127,14}
\definecolor{mplyellow}{RGB}{255,191,0}

\pgfmathsetmacro{\bandfactor}{1.0}

\begin{tikzpicture}
\begin{axis}[
  width=0.47\linewidth, height=0.3\linewidth,
  xmin=0.05, xmax=4, ymin=0, ymax=20,
  xtick distance=0.5,
  grid=both, 
  xlabel={Frequency (\si{\kilo\hertz})},
  ylabel={$\mathcal{D}$ (\si{\decibel})},
  legend style={at={(0.47,0.07)}, anchor=south west},
  legend cell align=left,
  tick label style={/pgf/number format/fixed},
]

\addlegendimage{line legend, thick, draw=mplblue}
\addlegendentry{$N=1,\ M=9$}

\addlegendimage{line legend, thick, dashed, draw=mplorange}
\addlegendentry{$N=2,\ M=9$}

\addlegendimage{line legend, thick, dotted, draw=mplyellow}
\addlegendentry{$N=3,\ M=9$}

\addplot[name path=hiA, draw=none]
  table[col sep=space,
        x expr=\thisrow{freq}/1000,
        y expr=\thisrow{mean_df} + \bandfactor*\thisrow{std_df}]
        {images/df_m999/data_m9_n1.csv};
\addplot[name path=loA, draw=none]
  table[col sep=space,
        x expr=\thisrow{freq}/1000,
        y expr=\thisrow{mean_df} - \bandfactor*\thisrow{std_df}]
        {images/df_m999/data_m9_n1.csv};
\addplot[forget plot, fill=mplblue, fill opacity=0.2]
  fill between[of=hiA and loA];
\addplot[thick, mplblue]
  table[col sep=space, x expr=\thisrow{freq}/1000, y=mean_df]
        {images/df_m999/data_m9_n1.csv};

\addplot[name path=hiB, draw=none]
  table[col sep=space,
        x expr=\thisrow{freq}/1000,
        y expr=\thisrow{mean_df} + \bandfactor*\thisrow{std_df}]
        {images/df_m999/data_m9_n2.csv};
\addplot[name path=loB, draw=none]
  table[col sep=space,
        x expr=\thisrow{freq}/1000,
        y expr=\thisrow{mean_df} - \bandfactor*\thisrow{std_df}]
        {images/df_m999/data_m9_n2.csv};
\addplot[forget plot, fill=mplorange, fill opacity=0.2]
  fill between[of=hiB and loB];
\addplot[thick, draw=mplorange, dashed]
  table[col sep=space, x expr=\thisrow{freq}/1000, y=mean_df]
        {images/df_m999/data_m9_n2.csv};

\addplot[name path=hiC, draw=none]
  table[col sep=space,
        x expr=\thisrow{freq}/1000,
        y expr=\thisrow{mean_df} + \bandfactor*\thisrow{std_df}]
        {images/df_m999/data_m9_n3.csv};
\addplot[name path=loC, draw=none]
  table[col sep=space,
        x expr=\thisrow{freq}/1000,
        y expr=\thisrow{mean_df} - \bandfactor*\thisrow{std_df}]
        {images/df_m999/data_m9_n3.csv};
\addplot[forget plot, fill=mplyellow, fill opacity=0.2]
  fill between[of=hiC and loC];
\addplot[thick, draw=mplyellow, dotted]
  table[col sep=space, x expr=\thisrow{freq}/1000, y=mean_df]
        {images/df_m999/data_m9_n3.csv};

\end{axis}
\end{tikzpicture}%
\label{fig:df_m999}}
\caption{Results of the Monte Carlo experiment in Sec.~\ref{sec:performance_increasing_order}. The top row shows the average (solid blue line) and target (dashed red line) hypercardioid beampatterns ($\theta_{\mathrm{s}} = \frac{\pi}{3}$) at \SI{1}{\kilo\hertz} for orders (a)~$N=1$, (b)~$N=2$, and (c)~$N=3$, synthesized using arbitrary arrays of $M=9$ microphones. The bottom row reports the corresponding average (d) white noise gain and (e) directivity factor as functions of frequency for the same target beampatterns and array configurations. Shaded regions denote one standard deviation around the mean. The higher number of microphones yields precise beampattern approximations with minimal variability across all orders.
Moreover, as the order increases the directivity factor grows while the white noise gain decreases, also reaching positive \SI{}{\decibel} values.}
\label{fig:montecarlo_m999}
\end{figure*}

\section{Validation}
\label{sec:04_validation}
This section validates and analyzes the proposed beamforming method using the metrics introduced in Sec.~\ref{sec:metrics} and Sec.~\ref{sec:metrics_new}, namely the beampattern, the white noise gain and the directivity factor.
Special attention is devoted to assessing the accuracy of the designed beampattern in approximating the target spatial response and whether this approximation remains consistent across frequencies.
The analysis considers target beampatterns of different shapes and orders, as well as various array geometries.

\subsection{Performance with Fixed Array Geometry}\label{sec:performance_fixed_geometry}
Let us first examine the performance of the proposed beamforming method when applied to a single fixed array geometry.
Results are shown in Fig.~\ref{fig:bp_frequency}. 
In particular, we consider the array configuration in Fig.~\ref{fig:geometry}, where $M=9$ first-order elements are randomly distributed in a two-dimensional plane. Every element is at least \SI{8}{\milli\meter} from its neighbors, and all lie in a circle of radius \SI{2}{\centi\meter}.
Such dimensions are compatible with modern small-size microphones \cite{zawawi2020review} and loudspeakers \cite{wang2021review}, for which differential beamforming is particularly intended. The directivity behavior of each element (concerning both the steering angle and the beam shape) is also drawn at random.

Figs.~\ref{fig:bp_freq_1} and \ref{fig:bp_freq_2} show the rendered beampatterns as a function of frequency in two specific cases: a first-order supercardioid and a fourth-order hypercardioid, respectively, both steered toward $\theta_{\mathrm{s}} = \frac{\pi}{3}$.
For reference, at the lowest frequency (\SI{50}{\hertz}) the target frequency-invariant beampattern is also shown.
It can be observed how the proposed method accurately reproduces the target response, maintaining strong frequency-invariant behavior across the entire band of interest, with consistent performance at higher orders.

\subsection{Performance with Random Array Geometries}
\label{sec:montecarlo}
To verify the validity of the method beyond a single configuration, we conduct three separate experiments, each consisting of $10000$ Monte Carlo trials covering a wide variety of array geometries and conditions. In each trial, elements are randomly placed within the same \SI{2}{\centi\meter}-radius aperture, with a minimum spacing of \SI{8}{\milli\meter}. Individual directivities are assigned by uniformly sampling the first-order beam shape coefficient $q_m$ and steering angle $\theta_m$ for each array element $m$.

\subsubsection{Robustness to Array Geometry Variations}\label{sec:robustness_geometry_variations}
The first experiment assesses the robustness of the proposed method's beampattern approximation to changes in array geometry.
In particular, we investigate the joint influence of beamformer order $N$ and the number of array elements $M$, with $M$ set to the theoretical minimum required to compute the minimum-norm solution in \eqref{eq:minimum_norm_solution}, namely $M=2N+1$.
Results are shown in Fig.~\ref{fig:montecarlo_m357}. Specifically, Figs.~\ref{fig:bp_fo_m3}, \ref{fig:bp_so_m5}, and \ref{fig:bp_to_m7} show the average rendered beampatterns at \SI{1}{\kilo\hertz} across all Monte Carlo simulations, together with the corresponding target patterns, for first-, second-, and third-order hypercardioids, respectively. The shaded areas denote one standard deviation around the mean spatial response.
Across all orders, and in particular when a higher number of array elements is considered, deviations from the target values remain minimal, with the most noticeable variations present at order $N=1$. The largest observed standard deviations range from \SI{8.8}{\decibel} at \SI{300}{\degree} for the first order, to \SI{7.4}{\decibel} at \SI{163}{\degree} for the third order, highlighting the robustness and consistency of the proposed design method against variations in array configuration.

White noise gain $\mathcal{W}[\mathbf{h}(\omega)]$ and directivity factor $\mathcal{D}[\mathbf{h}(\omega)]$ are shown in Figs.~\ref{fig:wng_m357} and \ref{fig:df_m357}, respectively, evaluated over the frequency range \SIrange{50}{4000}{\hertz}, with mean and standard deviation computed across the $10000$ Monte Carlo trials.
Consistent with traditional differential array theory \cite{benesty2016fundamentals}, $\mathcal{W}[\mathbf{h}(\omega)]$, shown in Fig.~\ref{fig:wng_m357}, decreases as the order increases, whereas $\mathcal{D}[\mathbf{h}(\omega)]$, in Fig.~\ref{fig:df_m357}, rises, reflecting the intrinsic trade-off between robustness to noise and spatial selectivity. 
Geometry variations only moderately affect either, with almost constant variability in white noise gain curves and rising variability in the high-frequency region of directivity factor curves.
Moreover, in Fig.~\ref{fig:wng_m357}, it can also be noticed that when considering the smallest number of microphones required to compute the beamforming filter at a specific order, the white noise gain drops below unity, resulting in degraded performance. Specifically, when $\mathcal{W}[\mathbf{h}(\omega)] < \SI{0}{\decibel}$, the power of the microphone self-noise is amplified \cite{wang2024design}.
Differential-array theory suggests that these limitations can be addressed by increasing the number of microphones in the array.

\subsubsection{Performance with Increasing Array Elements}\label{sec:performance_increasing_elements}
As a second experiment, we assess the performance of the proposed method when fixing  the target beampattern, while varying the number of microphones $M$.
Fig.~\ref{fig:montecarlo_m5913} presents the results of $10000$ Monte Carlo trials for rendering a second-order ($N=2$) supercardioid beampattern with $M=5, 9, 13$ array elements, under the arbitrary geometry conditions described in Sec.~\ref{sec:montecarlo}.
As expected, the beampattern approximations in Figs.~\ref{fig:bp_so_m5_v2}, \ref{fig:bp_so_m9_v2} and \ref{fig:bp_so_m13} improve as the number of microphones increases. The largest standard deviation occurs for the configuration with $M=5$ microphones with \SI{7.6}{\decibel} at \SI{166}{\degree}, decreasing to \SI{4.5}{\decibel} at \SI{314}{\degree} in the worst case for $M=13$. 
Such results verify that increasing the number of elements results in better performance.
The same behavior is also reflected in Fig.~\ref{fig:wng_m5913}, in which $\mathcal{W}[\mathbf{h}(\omega)]$ increases with the number of microphones, leaving $\mathcal{D}[\mathbf{h}(\omega)]$ (in Fig.~\ref{fig:df_m5913}) almost unaltered, as the target beampattern does not vary.

\subsubsection{Performance with Increasing Beamformer Order}\label{sec:performance_increasing_order}
As a final experiment, we evaluate the impact of the target beampattern order on the beamforming metrics.
To isolate this effect, the number of array elements is fixed at $M=9$, while the order varies ($N=1,2,3$).
Fig.~\ref{fig:montecarlo_m999} shows the results of $10000$ trials, based on the arbitrary array geometry described in Sec.~\ref{sec:montecarlo} and the target beampatterns in Sec.~\ref{sec:robustness_geometry_variations}.
As shown in Figs.~\ref{fig:bp_fo_m9}, \ref{fig:bp_so_m9}, and \ref{fig:bp_to_m9}, across all orders, the high number of microphones yields precise beampattern approximations with minimal variability, the worst standard deviation being \SI{6.9}{\decibel} at \SI{180}{\degree} for the first order.
As for the white noise gain and directivity factor, once again the results align with well-established results in differential array theory: increasing the order raises the directivity factor while lowering the white noise gain. Fig.~\ref{fig:wng_m999} shows that the variability of $\mathcal{W}[\mathbf{h}(\omega)]$ is significantly reduced compared to Figs.~\ref{fig:wng_m357} and \ref{fig:wng_m5913}, also reaching positive values. Similarly, Fig.~\ref{fig:df_m999} demonstrates a comparable reduction in the variability of $\mathcal{D}[\mathbf{h}(\omega)]$, further confirming the performance improvements.

\section{Conclusion}
\label{sec:05_conclusion}
In this work, we introduced a general framework for frequency-invariant differential beamforming on arbitrary planar arrays of first-order directional elements. By expressing the frequency-dependent directivity of each element via a circular harmonic decomposition and adopting a modal-matching approach, we decouple beampattern synthesis from both array geometry and element orientation. This extension naturally reduces to the traditional FIB-LSE method when all elements are omnidirectional, yet affords the flexibility to realize arbitrary-order directivity patterns with compact, arbitrarily configured arrays of first-order elements.

Extensive validation via both single-case examples and large-scale Monte Carlo trials demonstrates that the proposed method accurately approximates target beampatterns across wide bandwidths, maintaining a strong frequency invariance even at higher orders while achieving signal-to-noise-ratio gains in line with differential beamforming theory.

Beyond its immediate applications to DMAs and DLAs, the presented modal-matching extension provides a unified design tool for any planar array of first-order sources or sensors.
Future work will focus on extending the framework to arrays of arbitrary order elements, as well as experimental validation on hardware prototypes. Finally, the framework may be further generalized to 3D array configurations, opening new possibilities for compact, high-performance beamforming in cutting-edge acoustic and ultrasonic systems.
\appendix[Derivation of Harmonic coefficients]
In this appendix, we derive equation \eqref{eq:fourier_coefficients_derived}.
Starting from \eqref{eq:fourier_coefficients}, we expand the term $T_m(\theta)$, obtaining
\begin{equation}\label{eq:p_t_expanded}
    \begin{aligned}
        & \xi_{n,m} = \frac{(1-q_m)}{2\pi}\int_{-\pi}^{\pi}e^{jkr_m\cos{(\theta-\phi_m)}}e^{-jn(\theta-\theta_m)} \,d\theta \\
    &\quad +\frac{q_m}{2\pi}\int_{-\pi}^{\pi}\cos(\theta-\theta_m)e^{jkr_m\cos{(\theta-\phi_m)}}e^{-jn(\theta-\theta_m)}\,d\theta.
    \end{aligned}
\end{equation}
By imposing the change of variable $\beta = \theta - \phi_m - \frac{\pi}{2}$, \eqref{eq:p_t_expanded} can be expressed as \eqref{eq:appendix_coefficients}, in which $\varphi_m = \theta_m-\phi_m$.
\begin{figure*}[!hb]
\hrule
\vspace{5mm}
\begin{dmath}\label{eq:appendix_coefficients}
\begin{aligned}
    \xi_{n,m}
    = e^{jn\varphi_m}e^{-jn\frac{\pi}{2}}\left[ \frac{(1-q_m)}{2\pi} \int_{-\pi}^{\pi} e^{-j(kr_m\sin{(\beta)} + n\beta)}\,d\beta - \frac{q_m}{2\pi} \int_{-\pi}^{\pi} \sin{(\beta - \varphi_m)}e^{-j(kr_m\sin{(\beta)} + n\beta)}\,d\beta \right]
\end{aligned}
\end{dmath}
\end{figure*}
\begin{figure*}[!hb]
\begin{dmath}\label{eq:appendix_coefficients_rewritten}
\begin{aligned}
    & \xi_{n,m} = e^{jn\varphi_m}e^{-jn\frac{\pi}{2}}\left[ (1-q_m) (-1)^nJ_n(kr_m) - q_m \left(\cos{(\varphi_m)}j(-1)^nJ'_n(kr_m) + \sin{(\varphi_m)}\frac{n}{kr_m}(-1)^nJ_n(kr_m) \right)\right]
\end{aligned}
\end{dmath}
\end{figure*}

In order to make each step of the derivation clear, we split \eqref{eq:appendix_coefficients} into two parts: first we compute the contribution of its first integral, and then we address the second integral. The two parts are treated separately in the following subsections, yielding \eqref{eq:appendix_coefficients_rewritten}.
Finally, exploiting the definition of the derivative of the Bessel function
\begin{equation}\label{eq:derivative_bessel}
    J'_n(x) = \tfrac{1}{2}\bigg( J_{n-1}(x) - J_{n+1}(x) \bigg),
\end{equation} and the Bessel function property \cite{watson1922treatise}
\begin{equation}
    \tfrac{2n}{x}J_n(x) = \bigg( J_{n-1}(x) + J_{n+1}(x) \bigg),
\end{equation}
from \eqref{eq:appendix_coefficients_rewritten} we can derive \eqref{eq:fourier_coefficients_derived}.

\subsection{First Integral Term}
The first integral of \eqref{eq:appendix_coefficients}, namely
\begin{equation}
\begin{aligned}
    &\frac{1}{2\pi}\int_{-\pi}^{\pi}e^{-j(kr_m\sin(\beta) + n\beta)}\,d\beta,
\end{aligned}
\end{equation}
can be rewritten by exploiting both the Hansen-Bessel formula \cite{watson1922treatise}, i.e., 
\begin{equation}\label{eq:hansen_bessel}
    J_n(x) = \frac{1}{2\pi}\int_{-\pi}^{\pi}e^{j(n\theta-x\sin\theta)} \,d\theta,
\end{equation}
and the property
\begin{equation}\label{eq:bessel_der_p1}
    J_{-n}(x) = (-1)^nJ_n(x),
\end{equation}
yielding, in the end, 
\begin{equation}
\begin{aligned}
    &\frac{1}{2\pi}\int_{-\pi}^{\pi}e^{-j(kr_m\sin(\beta) + n\beta)}\,d\beta \\ &= \frac{1}{2\pi}\int_{-\pi}^{\pi}e^{j(-n\beta-kr_m\sin(\beta))}\,d\beta \\&= J_{-n}(kr_m)= (-1)^nJ_n(kr_m).
\end{aligned}
\end{equation}

\subsection{Second Integral Term}
Let us now focus on the second integral term of \eqref{eq:appendix_coefficients}, namely
\begin{equation}\label{eq:second_integral}
\begin{aligned}
    &\frac{1}{2\pi}\int_{-\pi}^{\pi}\sin(\beta-\varphi_m)e^{-j(kr_m\sin{(\beta)}+n\beta)}\,d\beta.
\end{aligned}
\end{equation}
By exploiting trigonometric identities, \eqref{eq:second_integral} can be expressed as
\begin{equation}\label{eq:double-swa}
\begin{aligned}
    &\frac{1}{2\pi}\int_{-\pi}^{\pi}\sin(\beta-\varphi_m)e^{-j(kr_m\sin{(\beta)}+n\beta)}\,d\beta \\
    & = \frac{1}{2\pi}\bigg( \int_{-\pi}^{\pi}\sin{(\beta)}\cos{(\varphi_m)}e^{-j(kr_m\sin{(\beta)}+n\beta)}\,d\beta \\
    &\quad - \int_{-\pi}^{\pi}\cos{(\beta)}\sin{(\varphi_m)} e^{-j(kr_m\sin{(\beta)}+n\beta)}\,d\beta \bigg) \\
    &= \frac{1}{2\pi}\bigg( \cos{(\varphi_m)}\int_{-\pi}^{\pi}\sin(\beta) e^{-j(kr_m\sin{(\beta)}+n\beta)}\,d\beta \\
    &\quad - \sin{(\varphi_m)}\int_{-\pi}^{\pi}\cos{(\beta)} e^{-j(kr_m\sin{(\beta)}+n\beta)}\,d\beta \bigg).
\end{aligned}
\end{equation}

The final form in \eqref{eq:double-swa} consists of two integral terms.
The first one can be rewritten as
\begin{equation}\label{eq:left_end_integral}
\begin{aligned}
    \frac{1}{2\pi} \int_{-\pi}^\pi \sin{(\beta)} e^{-j(kr_m\sin{(\beta)} + n\beta)} = j(-1)^nJ'_n(kr_m).
\end{aligned}
\end{equation}
This can be demonstrated by starting from \eqref{eq:derivative_bessel} and rewriting it exploiting \eqref{eq:hansen_bessel} as 
\begin{equation}\label{eq:left_end_integral_expanded}
\begin{aligned}
    & J'_n(kr_m) = \tfrac{1}{2} \bigg( J_{n-1}(kr_m) - J_{n+1}(kr_m) \bigg) \\
    & = \tfrac{1}{2} \bigg(
          \tfrac{1}{2\pi}
            \int_{-\pi}^{\pi}
              e^{j(n\beta - \beta - kr_m\sin(\beta))}
            \,d\beta 
        \\
    &\quad - \tfrac{1}{2\pi}
            \int_{-\pi}^{\pi}
              e^{j(n\beta + \beta - kr_m\sin(\beta))}
            \,d\beta
        \bigg) \\
    & = \tfrac{1}{2} \bigg(
          \tfrac{1}{2\pi}
            \int_{-\pi}^{\pi}
              e^{j(n\beta - kr_m\sin(\beta))}
              e^{-j\beta}
            \,d\beta
        \\
    &\quad - \tfrac{1}{2\pi}
            \int_{-\pi}^{\pi}
              e^{j(n\beta - kr_m\sin(\beta))}
              e^{j\beta}
            \,d\beta
        \bigg) \\
    & = \tfrac{1}{2} \bigg(
          \tfrac{1}{2\pi}
            \int_{-\pi}^{\pi}
              \cos(\beta)\,e^{j(n\beta - kr_m\sin(\beta))} \\
              &\quad - j\sin(\beta)\,e^{j(n\beta - kr_m\sin(\beta))}
            \,d\beta
        \\
    &\quad - \tfrac{1}{2\pi}
            \int_{-\pi}^{\pi}
              \cos(\beta)\,e^{j(n\beta - kr_m\sin(\beta))} \\
              &\quad + j\sin(\beta)\,e^{j(n\beta - kr_m\sin(\beta))}
            \,d\beta
        \bigg) \\
    & = -j \tfrac{1}{2\pi}
          \int_{-\pi}^{\pi}
            \sin(\beta)\,e^{j(n\beta - kr_m\sin(\beta))}
          \,d\beta.
\end{aligned}
\end{equation}
Moreover, from \eqref{eq:bessel_der_p1}, we can also demonstrate that $J'_{-n}(x) = (-1)^n J'_n(x)$, since
\begin{equation}
\begin{aligned}
    &\frac{\partial}{\partial x}J_{-n}(x) =
    (-1)^n \frac{\partial}{\partial x} J_n(x) =
    (-1)^n  J'_n(x).
\end{aligned}
\end{equation}
Therefore, by changing variable $\eta = -n$ in \eqref{eq:left_end_integral_expanded}, we get
\begin{equation}
\begin{aligned}
    &-j \frac{1}{2\pi} \int_{-\pi}^{\pi} \sin{(\beta)} e^{-j(\eta\beta + kr_m\sin{(\beta)})} \,d\beta \\
    &= J'_{-\eta}(kr_m) = (-1)^\eta J'_\eta(kr_m),
\end{aligned}
\end{equation}
thus demonstrating \eqref{eq:left_end_integral}.

Consider now the second integral in the final form of \eqref{eq:double-swa}. It can be rewritten by multiplying and dividing by $-jkr_m$ and applying once again \eqref{eq:hansen_bessel}, yielding
\begin{equation}\label{eq:right_end_integral_expanded}
\begin{aligned}
    &-\frac{1}{j2\pi kr_m} \int_{-\pi}^{\pi} -jkr_m\cos{(\beta)} e^{-j(kr_m\sin{(\beta)}+n\beta)} \,d\beta \\
    &= -\frac{1}{j2\pi kr_m} \int_{-\pi}^{\pi} -jkr_m\cos{(\beta)} e^{-j(kr_m\sin{(\beta)}+n\beta)} \\& \quad + jne^{-j(kr_m\sin{(\beta)}+n\beta)} -jn e^{-j(kr_m\sin{(\beta)}+n\beta)}\,d\beta\\
    &= -\frac{1}{j2\pi kr_m} \int_{-\pi}^{\pi} -j(kr_m\cos{(\beta)}+n) e^{-j(kr_m\sin{(\beta)}+n\beta)} \,d\beta \\& \quad -\frac{1}{j2\pi kr_m} \int_{-\pi}^{\pi} jn e^{-j(kr_m\sin{(\beta)}+n\beta)} \,d\beta \\
    & =-\frac{1}{j2\pi kr_m}\left[ e^{-j(kr_m\sin{(\beta)}+n\beta)}\Big|_{-\pi}^\pi \right] \\& \quad - \frac{n}{2\pi kr_m} \int_{-\pi}^{\pi} e^{-j(kr_m\sin{(\beta)}+n\beta)} \,d\beta \\& = - \frac{n}{kr_m} (-1)^n J_n(kr_m),
\end{aligned}
\end{equation}
for $n \in \{-N,\dots,N\}$.

\bibliographystyle{ieeetr}
\bibliography{refs}

\end{document}